\documentclass[hidelinks]{article}
\PassOptionsToPackage{numbers, compress}{natbib}
\usepackage[final]{neurips_data_2024}

\usepackage[utf8]{inputenc} 
\usepackage[T1]{fontenc}    
\usepackage{hyperref}       
\usepackage{url}            
\usepackage{booktabs}       
\usepackage{amsfonts}       
\usepackage{nicefrac}       
\usepackage{microtype}      
\usepackage{xcolor}         

\usepackage{graphicx}
\usepackage{enumerate}
\usepackage[shortlabels, inline]{enumitem}
\usepackage{multirow}
\usepackage{multicol}
\usepackage{balance}

\usepackage{ragged2e}

\usepackage{listings}
\usepackage{pifont}
\newcommand{\cmark}{\ding{51}} 
\newcommand{\xmark}{\ding{55}} 
\usepackage{tablefootnote}

\usepackage[english, status=draft, margin=false, inline=true]{fixme}
\usepackage{appendix}
\usepackage{fancyvrb}
\usepackage{tablefootnote}
\fxusetheme{color}
\FXRegisterAuthor{jf}{ajf}{\color{blue}JF}

\usepackage{array}
\usepackage{colortbl}
\usepackage{subcaption}
\usepackage{amsmath}
\usepackage{amssymb}
\usepackage{courier}
\usepackage{bbm}
\usepackage{adjustbox}
\usepackage{pifont} 
\usepackage{tabularx}

\newcommand{\rmspace}{\vspace{-2.2ex}}

\title{SM3-Text-to-Query: \underline{S}ynthetic \underline{M}ulti-\underline{M}odel \underline{M}edical Text-to-Query Benchmark}

\author{
    Sithursan Sivasubramaniam$^{*}$, 
    Cedric Osei-Akoto$^{*}$, 
    Yi Zhang, \\
    \textbf{Kurt Stockinger},
    \textbf{Jonathan Fürst$^{\dagger}$} \\
    Zurich University of Applied Sciences, Switzerland 
    \\
    \texttt{\{sivassit,oseiaced\}@students.zhaw.ch},
    \texttt{\{zhay,stog,fues\}@zhaw.ch} \\
}

\begin{document}

\maketitle
\renewcommand{\thefootnote}{\fnsymbol{footnote}}
\footnotetext{$*$ Equal contribution.}
\footnotetext{$\dagger$ Corresponding author.}

\begin{abstract}
Electronic health records (EHRs) are stored in various database systems with different database models on heterogeneous storage architectures, such as relational databases, document stores, or graph databases. These different database models have a big impact on query complexity and performance. While this has been a known fact in database research, its implications for the growing number of Text-to-Query systems have surprisingly not been investigated so far.
In this paper, we present SM3-Text-to-Query, the first multi-model medical Text-to-Query benchmark based on synthetic patient data from Synthea, following the SNOMED-CT taxonomy---a widely used knowledge graph ontology covering medical terminology. SM3-Text-to-Query provides data representations for relational databases (PostgreSQL), document stores (MongoDB), and graph databases (Neo4j and GraphDB (RDF)), allowing the evaluation across four popular query languages, namely SQL, MQL, Cypher, and SPARQL.
We systematically and manually develop 408 template questions, which we augment to construct a benchmark of 10K diverse natural language question/query pairs for these four query languages (40K pairs overall). On our dataset, we evaluate several common in-context-learning (ICL) approaches for a set of representative closed and open-source LLMs.
Our evaluation sheds light on the trade-offs between database models and query languages for different ICL strategies and LLMs. Last,
SM3-Text-to-Query is easily extendable to additional query languages or real, standard-based patient databases.
\end{abstract}

\section{Introduction}

The health sector is being increasingly digitalized, with data stored in
electronic health records (EHR)~\cite{menachemi2011benefits}.
In practice, those records can be kept in various
forms and systems: (i) traditional relational databases such as PostgreSQL or
Oracle implement the \emph{relational data model} where data is
organized into relations (tables) that are collections of tuples (rows). Users
access data through the declarative query language SQL; (ii) popular document
databases, such as MongoDB, model data as a collection of documents in the
\emph{document data model}, providing data access through specialized languages
such as the MongoDB Query Language (MQL); (iii) graph databases (triple stores)
model data as property graphs (e.g., Neo4j) or semantic RDF graphs (e.g.,
Ontotext GraphDB), providing interfaces through the query languages Cypher and SPARQL,
respectively. 

While the relational model and the SQL query language are still the primary choice
for EHRs~\cite{lee2022ehrsql}, there has been an increased interest in document and graph database models due to their schema flexibility and natural capacity to interconnect data sources across data silos~\cite{solmaz2022enabling, furst2023versamatch, cheng2024interactive}. E.g., \textsc{AICCELERATE}, a recent large-scale European pilot project on digital hospitals, uses a graph model to connect various data sources, including traditional EHRs, other hospital data, wearables, and IoT devices~\cite{aiccelerate}. Moreover, RDF graph databases that use SPARQL as the primary query language are widely used in life sciences, with medicine rapidly catching up~\cite{sima2022bio}.

\begin{figure}[ht]
    \centering
    \includegraphics[width=1.0\textwidth]{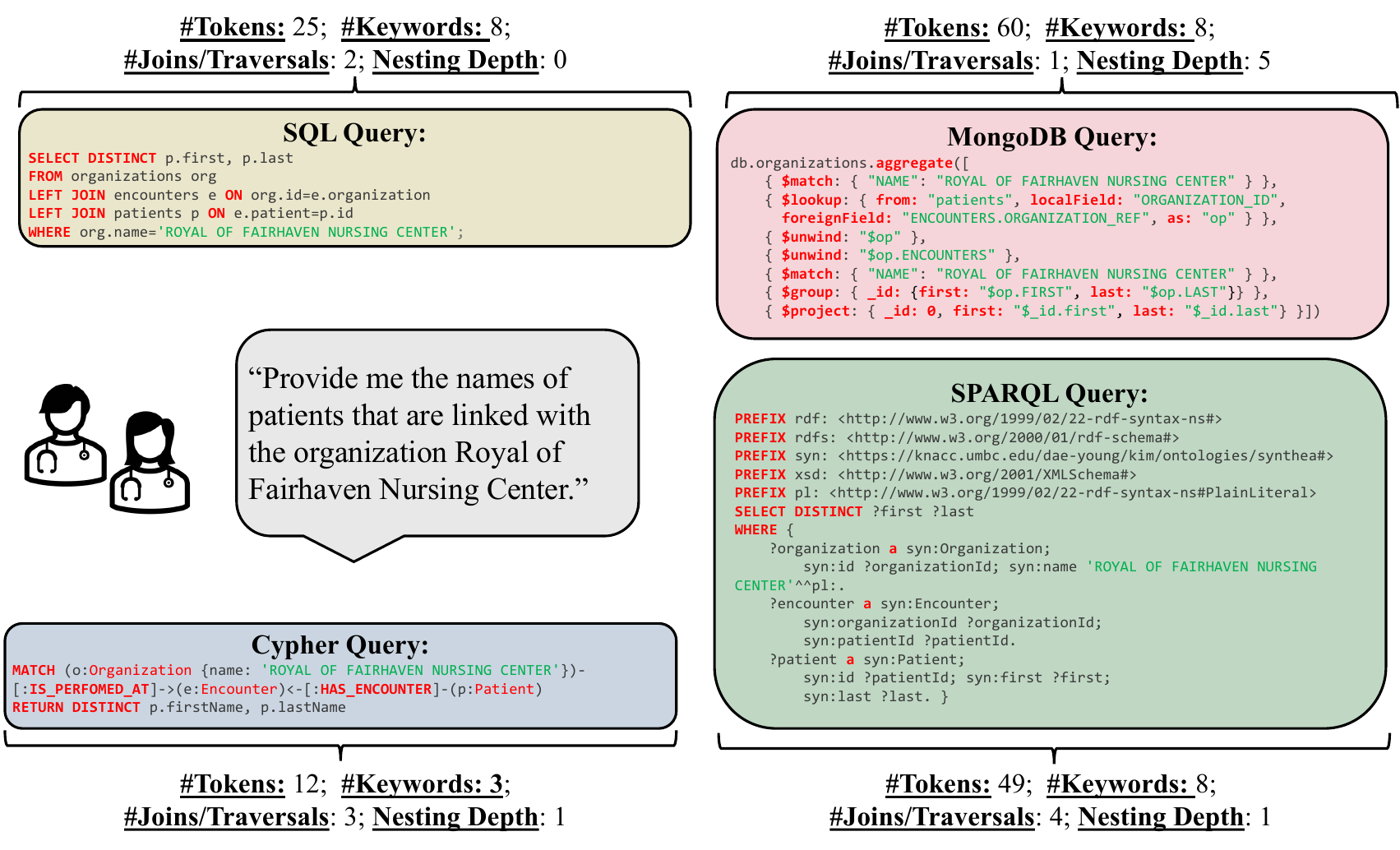}
    \caption{Differences across query languages and database systems for the same user request. }
    \label{fig:querydifference}
    \rmspace
\end{figure}

The choice of database and the underlying core data model (relational, document,
graph) has a large impact on read/write performance and query complexity.
For example, the graph model naturally represents many-to-many relationships, such as connections between patients, doctors, treatments, and medical conditions,
whereas relational databases require potentially expensive join operations and complex queries. Document databases have only rudimentary
support for many-to-many relationships and aim at scenarios where data is
not highly interconnected and stored in collections of documents with a flexible schema~\cite{kleppmann2017designing}.
Figure~\ref{fig:querydifference} shows an example of a single user question together with the corresponding query statements in four query languages. 
While these differences have been a known fact in database research and industry, \emph{its
implications for the growing number of Text-to-Query systems have surprisingly
not been investigated so far.}

Text-to-Query systems have seen a recent growth in the number of developed
methods and new high scores, mainly due to the transformer architecture and
advances in Large Language Models (LLMs)~\cite{katsogiannis2023survey}. The idea is compelling: Instead of
writing database queries, users express their intends in natural language and
a Text-to-Query system with access to the underlying database translates them
into valid query statements (e.g., SQL or SPARQL). This is especially relevant in
a digital health context in which the domain experts (e.g., nurses, doctors, or admin
staff) cannot be expected to write complex queries and currently rely on
pre-defined rule conversion systems that limit the potential questions that can
be asked~\cite{lee2022ehrsql}.

Existing Text-to-Query datasets and benchmarks have usually been focusing on
single database models and query languages. E.g., Spider~\cite{yu2018spider},
WikiSQL~\cite{zhong2017seq2sql}, BIRD~\cite{li2024can}, ScienceBenchmark~\cite{zhang2023sciencebenchmark}, Statbot.swiss~\cite{noor2024} and
EHRSQL~\cite{lee2022ehrsql} (Text-to-SQL) and LC-Quad 2~\cite{dubey2019lc} and Spider4SPARQL~\cite{kosten2023spider4sparql},  for
Text-to-SPARQL, also called Knowledge Graph Question Answering (KGQA). Only
recently, in FootballDB~\cite{furst2024evaluating}, have there been initial attempts to evaluate the data model impact inside a single database system (with different schemas for the same
data). Further, there has been initial works across two database models and query
languages. E.g., a recent comparison of SQL and
SPARQL based on the MIMICSQL dataset~\cite{wang2020text} finds a 34-point accuracy difference~\cite{park2021knowledge}.

However, \textit{no existing work in and outside the medical domain enables the evaluation of Text-to-Query
across multiple core database models and query languages}. This is the goal of our dataset and benchmark. In this paper, we present
\emph{SM3-Text-to-Query, a first Multi-Model Medical Text-to-Query Benchmark
based on synthetic patient data}. SM3-Text-to-Query provides the following key features:

\begin{itemize}
    \item \textbf{Standard-based and privacy-preserving.} SM3-Text-to-Query has
    been carefully constructed from Synthea~\cite{walonoski2018synthea},
    a synthetic patient data simulator. Thus, there are no privacy implications
    in the published dataset.  In our data schemas, we follow the SNOMED-CT
    taxonomy~\cite{SNOMED-International:2024aa}, a commonly used medical knowledge graph ontology that is widely applied across
    institutions and countries, enabling interoperability of health data.
    \emph{This ensures the wide impact of the benchmark on real-world health
    use cases.}
    
    \item \textbf{Three database models---four database systems and query languages.} Involving
    database experts in the process, we design four data representations (schemas) in
    PostgreSQL, MongoDB, Neo4j, and RDF (GraphDB), and create transformations for
    them from the Synthea output. These databases represent three core database models: relational, document, and graph model.
    The databases with \emph{the same data content} can be accessed through four
    different query languages: SQL, MQL, Cypher, and SPARQL. SM3-Text-to-Query is, to our knowledge, the first benchmark with these features.
    \emph{Our chosen database systems and query languages constitute a wide
    representation of the most popular systems according to DB-Engines
    Ranking~\cite{db-ranking2024}.}
    
    \item \textbf{Systematic and expandable question generation.} We systematically and manually create a set of $408$ template questions covering the major entities and properties of the Synthea data. These template questions
    are then automatically enhanced and enriched to result in a benchmark set of 10K natural language/query pairs for each query language. The enrichment is performed via parameterizable sampling methods to retrieve, for instance, patient names or allergies from the underlying
    database. Our method is easily extendable through the addition of new
    templates (e.g., different languages, different questions, paraphrasing)
    or through plugging in real patient databases modeled according to
    SNOMED-CT.
    \emph{This ensures the benchmark is future-proof and can be adjusted to other use scenarios.}
    
\end{itemize}

\section{Related Work}

In this section, we review related works for Text-to-Query systems with a focus on (i) medical data and (ii) generally relevant benchmarks.

\textbf{Medical focus.}
MIMICSQL~\cite{wang2020text} is derived from the MIMIC-III database~\cite{johnson2016mimic}, containing 10K unique questions tailored to medical quality assurance tasks. To avoid potential limitations, such as fixed question structures, MIMICSQL underwent a filtering and paraphrasing process performed by expert freelancers.
MIMIC-SPARQL~\cite{park2021knowledge} builds on the framework of MIMICSQL~\cite{wang2020text} and customizes its question templates to query a modified schema with SPARQL. With a similar structure to MIMICSQL, it provides 10K unique questions tailored to medical QA tasks.
Last, EHRSQL~\cite{lee2022ehrsql} provides a benchmark for text-to-SQL tasks with a focus on electronic health records (EHR). It is based on MIMIC-III and eICU databases, while the 230 question templates are derived from user surveys. Based on these 230 templates, EHRSQL generates 24K questions/query pairs for SQL.

\textbf{General benchmarks.}
The WikiSQL~\cite{zhong2017seq2sql} dataset is a well-known general Text-to-SQL benchmark that comprises over 80K text/SQL pairs. What makes this dataset noteworthy is the wide distribution of queries over 24,241 tables.
Spider~\cite{yu2018spider} is considered one of the most popular cross-domain text-to-SQL datasets and consists of 10,181 questions with 5,693 unique SQL queries on 200 databases. 
KaggleDBQA~\cite{lee-etal-2021-kaggledbqa} builds on large-scale datasets such as Spider and WikiSQL to provide a cross-domain dataset with domain-specific data types.
BIRD~\cite{li2024can} is a comprehensive resource for question answering (QA) that includes 12,751 unique questions from various repositories such as Kaggle, CTU Prague, and Open Tables and covers 37 subject areas. BIRD targets real-world applications by including complex examples from 95 large databases totaling 33.4 GB.
%
ScienceBenchmark~\cite{zhang2023sciencebenchmark} presents three real-world, domain-specific text-to-SQL datasets. In comparison to other datasets, it reflects the high importance of domain-specific benchmark datasets for real-world text-to-SQL tasks.
Last, FootballDB~\cite{furst2024evaluating} investigates different database schemas and their impact on Text-to-SQL systems inside a single database. The questions are derived from a real deployment with end users.

Compared to these works, the main novelty of SM3-Text-to-Query is that it is, to the best of our knowledge, the first dataset and benchmark that allows for the evaluation of Text-to-Query systems across three core database models (relational, graph, document) and four query languages (SQL, SPARQL, Cypher, MQL). FootballDB~\cite{furst2024evaluating} is comparable in terms of analyzing the schema dependency of Text-to-SQL systems inside a single database. However, it only targets SQL, a single database model (relational), and query language (SQL). For our template-based approach, we take inspiration from EHRSQL~\cite{lee2022ehrsql}.
We complement their idea with synthetic data generation following a widely used medical standard (SNOMED~\cite{SNOMED-International:2024aa}). This makes our benchmark relevant to digital health scenarios worldwide, where databases follow the SNOMED medical naming taxonomy.

\section{SM3-Text-to-Query Benchmark Construction}

\begin{figure}[ht]
    \centering
    \includegraphics[width=1.0\textwidth]{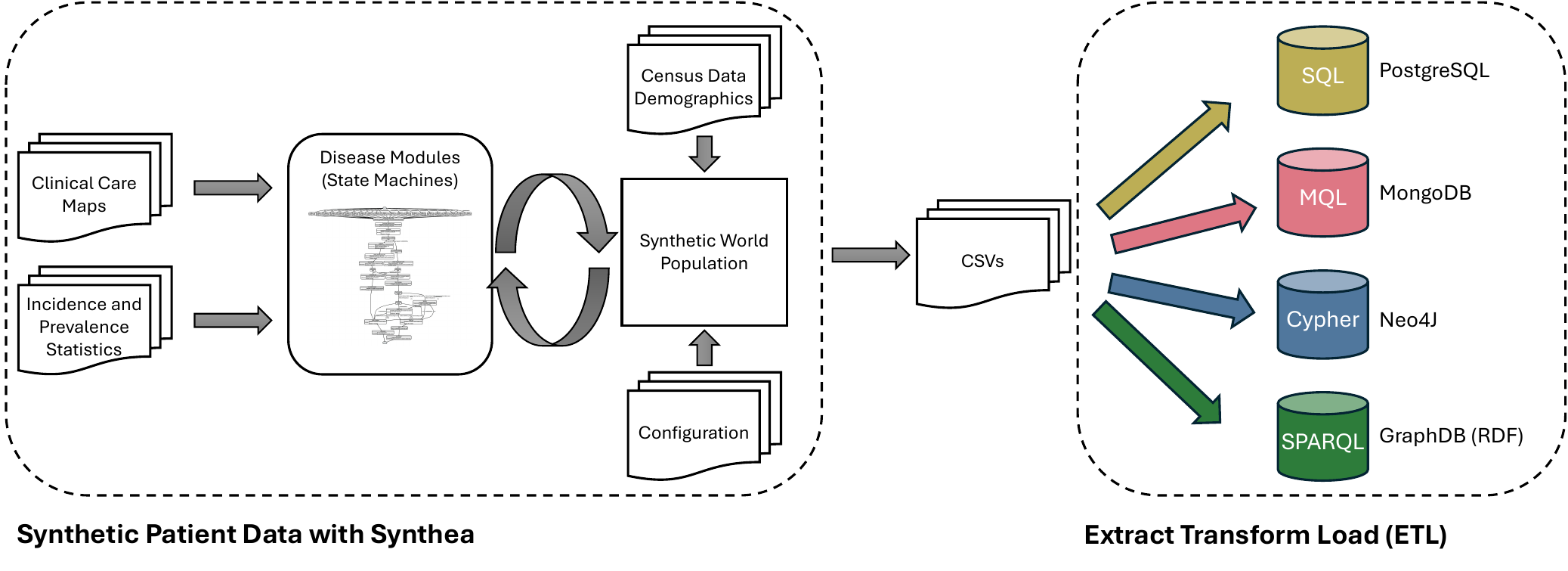}
    \caption{\textbf{Database construction from Synthetic Patient Data}. Synthea uses clinical care maps and statistics to build models of disease progression and treatment, encoded as state transition machines. The synthetic world population is seeded with census data demographics and configuration options, enabling the creation of realistic patient data in 18 classes, which we export as CSVs. We implement custom Extract Transform and Load (ETL) pipelines to transform these CSVs to four database systems and models.}
    \label{fig:database-construction}
        \rmspace
\end{figure}

Our SM3-Text-to-Query benchmark construction consists of two main steps. First, we construct the database based on synthetic medical data in four data models (Figure~\ref{fig:database-construction}). Second, we implement a template-based text/query-pair construction approach. The dataset was created over a period of more than a year in the context of a project with health professionals from two university hospitals, medical doctors, nurses, data scientists, and computer scientists who have been working in the respective fields for 5+ years. We constructed the question templates to cover all SNOMED CT entities. The database queries were mainly written by two undergraduates and one PhD student with a computer science and database background and verified by two faculty members. 

\subsection{Database Construction}
\label{sub:dataset_construction}

\subsubsection{Synthetic Patient Data with Synthea}

As much of this data only reaches its full potential when it is interoperable across organizations, such as hospitals, insurance providers, and specialists, there has been a move toward standardization in healthcare. Here, SNOMED Clinical Terms (CT) is considered to be the most comprehensive, multilingual clinical healthcare terminology in the world~\cite{benson2021principles}.
While there exists a range of clinical EHR datasets such as MIMIC-III~\cite{johnson2016mimic} and eICU~\cite{pollard2018eicu}, they are based on de-identified data from single countries (eICU) or even just single medical centers (MIMIC-III) and do not follow the SNOMED CT taxonomy.

To construct the SM3-Text-to-Query benchmark, we, therefore, choose to build on synthetic but standardized data that we generate through Synthea~\cite{walonoski2018synthea}. 
Synthea is an open-source, synthetic patient generator that models the medical history of synthetic patients and their electronic health records while being
compatible with SNOMED CT. Its generated data includes 18 classes representing various aspects of healthcare, such as allergies, care plans, and medications, and is available in CSV format, FHIR, C-CDA, and CPCDS formats~\cite{synthea_github}.
Synthea employs top-down and bottom-up approaches to generate structured synthetic EHRs throughout a patient's life. E.g., the simulation incorporates models for the top ten reasons for primary care visits and chronic conditions responsible for the most years of life lost in the United States, based on US Census Bureau, Centers for Disease Control and Prevention, and National Institutes of Health reports. Further, international populations can be simulated through provided metadata and configuration files for currently 25 countries~\cite{synthea_international}.

\subsubsection{Data Transformation to different Databases}

Based on the Synthea output, we define data schemas/ontologies and implement Extract, Transform and Load (ETL) pipelines for our chosen databases: PostgreSQL (SQL), MongoDB (MQL), Neo4j (Cypher), GraphDB (RDF).
For PostgreSQL, we define appropriate data types and consistency constraints (mainly primary and foreign keys). MongoDB, as a document database, does not enforce a strict schema as relational databases. Here, we define a JSON schema following a tree structure with four top-level collections (patients, organizations, providers, payers) and the remaining entities being embedded in these collections. We connect documents across collections through ID references (\verb|$lookup| operator as equivalent to Join). 
For Neo4J, we implement ETL by following the guidelines provided by Neo4j Solutions~\cite{ingest_neo4j}, adapting it due to missing classes and connections in the original configuration files.
For RDF, we extend Synthea-RDF~\cite{synthea_github} to cover all Synthea attributes and automate the conversion of data from CSV files into RDF as Terse RDF Triple Language (TTL).
All four data models can be found in Appendix~\ref{app:db-schemas} and in the supplement material.

\subsection{Text/Query-Pairs Construction}

To construct a dataset of text/query-pairs, we follow the established template-based approach~\cite{finegan2018improving, wang2020text, park2021knowledge, lee2022ehrsql} in which a set of template questions is augmented with values to scale the dataset without extensive manual efforts. Together with the standard-based Synthea data (SNOMED CT taxonomy), our generation process has the following advantages: 
\begin{itemize}

    \item \textbf{Coverage and diversity through Synthea data generation.} Through the use of templates, the benchmark dataset can be automatically adjusted to different Synthea datasets (e.g., for different patient populations). Each template is tagged with the relevant Synthea entities (e.g., patient, claim). Based on this structure, our method allows for the construction of
    datasets that only cover a subset of entities (e.g., only focusing on an insurance database).
    \item \textbf{Reduced bias in machine learning methods of the task.} By filling the query templates with parameterizable values, various biases of text-to-query methods can be exploited. Thus, we can avoid the respective LLMs overfitting. 
    \item \textbf{Standardized evaluation over different systems.} The creation of standardized templates is possible through the implementation of the SNOMED terminology in the Synthea dataset. The benchmark dataset leverages SNOMED attributes, i.e., standardized medical terminology, for the evaluation of queries over different systems and database models. The same template questions can easily be combined with real-patient data following SNOMED medical terminology.
    
\end{itemize}

Overall, we create 408 template questions (see supplement material for a full list and Appendix~\ref{app:template-example} for an example) in a structured way that is guided by the goal to cover all 18 entity types created by Synthea. The template questions include WH\footnote{Who, What, Where, When, and Why.} and non-WH questions, factual questions, linking questions, summarization questions, and cause-effect questions. We tag each template question with its related entities and question type.

Last, following previous work~\cite{lee2022ehrsql}, we define 10 non-answerable medical and 5 non-medical questions. These are questions that cannot be answered from data stored in the respective databases but would require additional information. For instance, \textit{What is the marital status of patient Max Müller?}

For each question template, we manually develop the corresponding query in SQL, SPARQL, Cypher and MQL.
The queries are then verified by a second expert for correctness. 
For scaling the template questions, we augment them by automatically inserting values such as IDs, descriptions of diseases, and patient names queried from the database. This data augmentation step is fully configurable and can be used to generate enriched and linguistically diverse text/query pairs for arbitrary Synthea databases and in-the-wild databases following the SNOMED CT standard.

\section{Dataset Analysis and Comparison}
\label{sec:dataset}

\begin{figure}[ht]
    \centering
    \begin{minipage}{0.49\textwidth}
        \centering
    \includegraphics[width=1.0\textwidth]{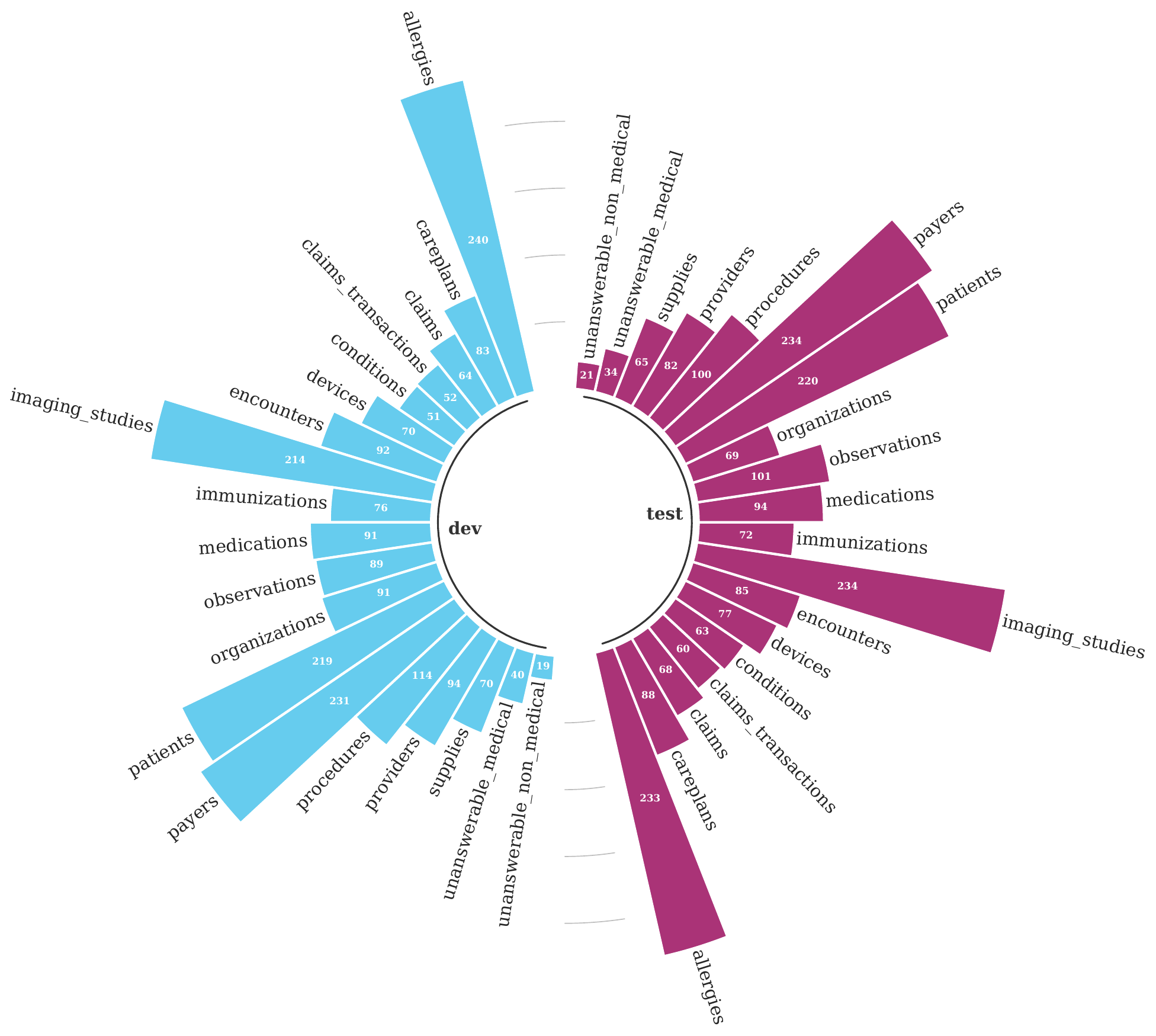}
    \end{minipage}\hfill
    \begin{minipage}{0.49\textwidth}
        \centering
    \includegraphics[width=1.0\textwidth]{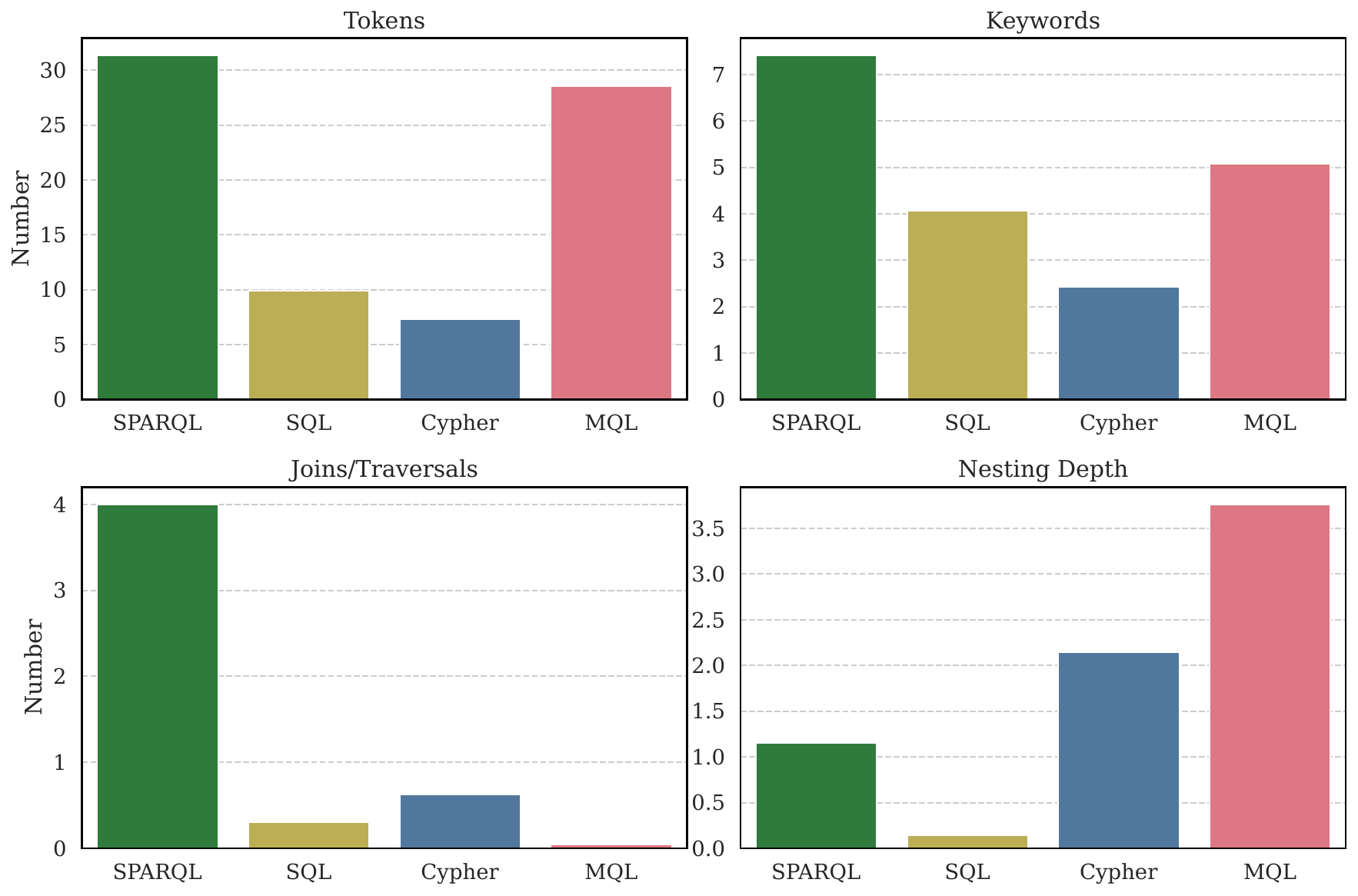}
    \end{minipage}
    \caption{Dataset distribution for dev and train (left); Query statistics for query languages (right).}
    \label{fig:dataset_images}
        \rmspace
\end{figure}

\textbf{Question and Query Statistics.}
For SM3-Text-to-Query, we use our method described in Section~\ref{sub:dataset_construction} to construct a synthetic multi-modal dataset of 10K text/query pairs for each of the four query languages (resulting in 40K individual samples). The dataset is based on the default Synthea configuration with medical records of 100 living and 10 diseased patients.
We split the data into 6K train, 2K dev, and 2K test with stratified sampling where the strata are the entity types that the questions are tagged with.
Figure~\ref{fig:dataset_images} summarizes dev and test, with the distribution of different types of user questions on the left. While our template questions cover all SNOMED entities, there exist more templates for allergies, imaging studies, patients, and payers, which is why they are over-represented in the augmented dataset.

While all queries across query languages return the same information for a single user question, they vary greatly in their characteristics (Figure~\ref{fig:dataset_images}, right).
For instance, SPARQL queries have the largest number of tokens and keywords, followed by MongoDB's MQL queries and SQL queries, while Cypher queries are the most compact ones.
We also count the number of joins/traversals and the nesting depth (based on opening/closing characters) of the queries, an established complexity measure in programming language research~\cite{alrasheed2014understanding}. Here, MQL requires notably the fewest joins (\verb|$lookup| operators) while exhibiting the highest nesting depth (due to the embedded collection structure in its schema).

\textbf{Comparison to other datasets.}
Table~\ref{tab:benchmark-comparison} summarizes SM3-Text-to-Query with respect to a selection of relevant datasets and benchmarks. SM3 has a similar number of example questions compared to other datasets, while the number of corresponding queries is substantially higher due to the translation to four query languages. Qualitatively, \emph{SM3 is the only dataset that enables an evaluation across four different query languages, which is not only unique for medical data but does currently not exist for any domain.}
Last, SM3 is standard-based (SNOMED), making it compatible with standard-based real health databases and extensible through our template-based approach.

\begin{table}[ht]
  \caption{A comparison between \textsc{SM3} and other relevant Text-to-Query datasets and benchmarks (NA-Qs: non-answerable questions; Standard: standard-based, e.g., SNOMED; Template: Template-based, which allows the dataset to be easily extended).} 
  \label{table:overview_Quantitative}
  \centering
  \resizebox{1.0\hsize}{!}{\begin{tabular}{c|cccccccccc c c c c}
    \toprule
    \textbf{Dataset} & 
    \textbf{Data source} & 
    \textbf{\# Questions} & 
    \textbf{\# Queries} & 
    \textbf{NA-Qs} & 
    \textbf{Medical} & 
    \textbf{Standard} & 
    \textbf{Template} & 
    \textbf{SQL} & 
    \textbf{MQL} & 
    \textbf{SPARQL} & 
    \textbf{Cypher} \\
    \midrule
    \textsc{MIMICSQL} \citep{wang2020text}  & MIMIC-III~\cite{johnson2016mimic} & 10,000 & 10,000
    & \textcolor{red}{\xmark}
    & \textcolor{teal}{\cmark}
    & \textcolor{red}{\xmark}
    & \textcolor{teal}{\cmark}
    & \textcolor{teal}{\cmark}
    & \textcolor{red}{\xmark}
    & \textcolor{red}{\xmark}
    & \textcolor{red}{\xmark} \\
    \textsc{MIMIC-SPARQL} \citep{park2021knowledge} & MIMIC-III~\cite{johnson2016mimic} & 10,000 & 10,000
    & \textcolor{red}{\xmark}
    & \textcolor{teal}{\cmark}
    & \textcolor{red}{\xmark}
    & \textcolor{teal}{\cmark}
    & \textcolor{red}{\xmark}
    & \textcolor{red}{\xmark}
    & \textcolor{teal}{\cmark}
    & \textcolor{red}{\xmark} \\
    \textsc{EHRSQL}~\citep{lee2022ehrsql} & MIMIC-III~\cite{johnson2016mimic}, eICU~\cite{pollard2018eicu} & 24,000 & 24,000
    & \textcolor{teal}{\cmark}
    & \textcolor{teal}{\cmark}
    & \textcolor{red}{\xmark}
    & \textcolor{teal}{\cmark} 
    & \textcolor{teal}{\cmark}
    & \textcolor{red}{\xmark}
    & \textcolor{red}{\xmark}
    & \textcolor{red}{\xmark} \\
    \textsc{BIRD} \citep{li2024can} & Kaggle, CTU Prague, Web & 12,751 & 12,751
    & \textcolor{red}{\xmark}
    & \textcolor{red}{\xmark}
    & \textcolor{red}{\xmark}
    & \textcolor{red}{\xmark} 
    & \textcolor{teal}{\cmark}
    & \textcolor{red}{\xmark}
    & \textcolor{red}{\xmark}
    & \textcolor{red}{\xmark} \\
    \midrule
    \textsc{\textbf{SM3 Text-to-Query}} & Synthea~\cite{synthea_international} & 10,000 & 40,000
    & \textcolor{teal}{\cmark}
    & \textcolor{teal}{\cmark}
    & \textcolor{teal}{\cmark}
    & \textcolor{teal}{\cmark} 
    & \textcolor{teal}{\cmark}
    & \textcolor{teal}{\cmark}
    & \textcolor{teal}{\cmark}
    & \textcolor{teal}{\cmark}\\
    \bottomrule
  \end{tabular}}
\label{tab:benchmark-comparison}
\end{table}

\textbf{Comparison of query complexity.} We also compare SM3 query complexity against the two common medical Text-to-SQL datasets EHRSQL~\cite{lee2022ehrsql} and MIMICSQL~\citep{wang2020text} (Table~\ref{tab:query-complexity}).
EHRSQL and MIMICSQL feature more complex SQL queries, such as temporal ones. However, the strength of our dataset lies in its cross-model aspect, which includes Cypher, MQL, and SPARQL. Our MQL and SPARQL queries have similar complexity in terms of token count. SPARQL even includes more joins and traversals than existing benchmarks, while MQL has more nesting.

\begin{table}[ht]
\caption{SM3-Text-to-Query query statistic compared to recent medical Text-to-SQL datasets.}
\label{tab:query-complexity}
\centering
\resizebox{0.75\hsize}{!}{
\begin{tabular}{lrrrrrr}
\toprule
& \textbf{EHRSQL} & \textbf{MIMICSQL} & \textbf{SM3\textsubscript{Cypher}} & \textbf{SM3\textsubscript{MQL}} & \textbf{SM3\textsubscript{SPARQL}} & \textbf{SM3\textsubscript{SQL}} \\
\midrule
\textbf{Tokens} & \textbf{32.63} & 19.45 & 7.28 & 28.56 & 31.35 & 9.93 \\
\textbf{Keywords} & \textbf{11.74} & 6.38 & 2.43 & 5.08 & 7.41 & 4.06 \\
\textbf{Joins/Traversals} & 0.33 & 0.64 & 0.62 & 0.04 & \textbf{4.00} & 0.31 \\
\textbf{Nesting Depth} & 1.37 & 0.77 & 2.14 & \textbf{3.77} & 1.15 & 0.15 \\
\bottomrule
\end{tabular}}
\end{table}

\section{Baseline Experimental Evaluation}
\label{sec:evaluation}

The goal of our experiments is to evaluate how well large language models (LLMs) perform in translating natural language questions into four different query languages provided by our novel benchmark. We restrict ourselves to LLMs as these models dominate the leader boards of popular Text-to-Query benchmarks such as Spider~\cite{yu2018spider}.

\subsection{Experimental Setup}
\label{sub:experimental-setup}

For our baseline experiments, we select a set of four common open and closed-sourced LLMs and implement the same in-context learning (ICL) prompting strategies across our four query languages. Here, we follow standard practices in terms of task instruction formulation and the inclusion of schema/ontology information as well as few-shot examples. We take inspiration from~\citeauthor{nan-etal-2023-enhancing} who propose general design strategies for enhancing Text-to-SQL capabilities in LLMs~\cite{nan-etal-2023-enhancing}, widely adopted in recent applications. Further, \citeauthor{chang2023prompt} demonstrate the importance of including database schema and content~\cite{chang2023prompt}. \citeauthor{liu2024episql} suggest using notes and annotations to mitigate unexpected behaviors of LLMs~\cite{liu2024episql}, improving  accuracy. Drawing upon these strategies, we developed various prompt templates tailored to the requirements of our experimental settings. The detailed templates applied in the experiments are listed in Appendix~\ref{sec:appendix_experimental_details}.

As open-source LLMs, we select Meta Llama3-8b and Llama3-70b (instruction-tuned variants)~\cite{llama3modelcard}. As closed-source LLMs, we select Google Gemini 1.0 Pro~\cite{team2023gemini} and OpenAI GPT-3.5-turbo-0125~\cite{openai-chatgpt}. The closed-sourced models are run via their respective APIs. We run Llama3-8b locally on a single NVIDIA A100 GPU. For Llama3-70b, we use the cloud-hosted model provided by Groq~\cite{groq}. 

We define three prompts with schema information (\emph{w/schema 0-shot}, \emph{w/schema 1-shot} and \emph{w/schema 5-shot}) and two prompts without schema information (\emph{w/o schema 1-shot} and \emph{w/o schema 5-shot}). Due to the size of the ontology for SPARQL, we implement a smaller version in which the ontology is summarized as a JSON object with all contained classes, objects, and data properties. Likewise, for MQL and Cypher, we provide the encoded MongoDB document schema and Neo4J graph schema, respectively.
The shots are selected using stratified random sampling by considering the categories of the original question templates to ensure a diverse selection. For the prompts that include examples (1-shot/ 5-shot), we perform five runs with different sampling (for further details, see Appendix~\ref{sec:appendix_experimental_details}).

\paragraph{Execution Accuracy (EA).}
We apply \emph{exact execution accuracy} (EA), also known as result matching~\cite{kim2020natural} as our main accuracy metric.
EA denotes the fraction of questions within the evaluation set, where the outcomes of both the predicted and ground-truth queries yield identical results relative to the total number of queries.
Given two sets, i.e., the reference set $R$, produced by the execution of the $n$ ground-truth SQL queries $Y_n$, and the corresponding result set denoted as $\hat{R}$ obtained from the execution of the predicted SQL queries $\hat{Y}_n$, EX can be computed by Equation~\ref{eq:ex}.

\begin{equation}
\label{eq:ex}
    EA = \frac{\sum_{n=1}^{N}{\mathbb{I}(r_n, \hat{r}_n)}}{N}
\end{equation}
where $r_n \in R_n$, $\hat{r}_n \in \hat{R}_n$, and $\mathbb{I}$ is the indicator function, defined as:
\begin{equation}
\label{eq:indicator_function}
    \mathbb{I}(r_n, \hat{r}_n) = \begin{cases}
      1 & \text{if $r_n = \hat{r}_n$}\\
      0 & \text{else}\\
    \end{cases}       
\end{equation}

\subsection{Text-to-Query Accuracy}
\label{subs:accuracy_overall}

We first evaluate Text-to-Query accuracy for the different prompting strategies, LLMs, and query languages. Table~\ref{tab:ea_test}  depicts the Execution Accuracy (EA) without schema information (two left-most columns), while the three right-most columns contain results for experiments with schema. The \textpm\ represents the standard deviation. Numbers are in \%.
We observe the following four key insights:

\begin{itemize}[leftmargin=*]

    \item \emph{Schema information helps for all query languages, but not equally.} As expected, the accuracy of w/ schema experiments is higher than that of their w/o schema counterparts. Especially with 1-shot, the models cannot generate correct queries in the majority of cases.
    However, surprisingly, schema information has a much larger impact on the performance for SQL, Cypher, and MQL (more than doubles the performance for 5-shot compared to w/o schema) than for SPARQL (only slightly higher or equal performance).
    This indicates that LLMs may have encountered the SNOMED CT ontology and vocabulary during their pre-training phase, as these are standardized and widely published on the web, whereas the specific schemas for SQL, Cypher, and MQL databases are private to each implementation and thus novel to the model, making explicit schema information more crucial for these query languages.

    \item \emph{Adding examples improves accuracy through ICL for all LLMs and query languages, however, the rate of improvement varies greatly across query languages.} For SQL---the most popular query language---the larger LLMs already achieve $\approx40$\%  (w/schema 0-shot) and only improve by $\approx10$ points with 5-shots ($\approx25\% $ relative improvement). For the ``more exotic'' query languages (SPARQL, MQL, and partly Cypher), LLMs are often unable to generate a correct query with only the schema information. E.g., for SPARQL, 0-shot is $<4\%$, while 5-shot can reach up to $30\%$ (10-fold relative improvement). This indicates again that the model is already proficient in the SQL query language, whereas for SPARQL (and to a smaller extent Cypher and MQL), the model is truly benefiting from ICL by learning how to formulate more correct queries from the provided fixed few-shot examples (even though the examples might not directly be related to the asked question).

    \item \emph{LLMs exhibit mostly consistent performance patterns across query languages.} Observing w/schema 5-shot results, Llama3-70b achieves the best results for all query languages. GPT-3.5 and Llama3-70b share the 2nd and 3rd place, while the smallest LLM Llama3-8b achieves always the lowest accuracy.    
    Further, some results show a large standard deviation, indicating that the different few-shot example compositions for each run have a large performance impact. To further investigate the impact of few-shot sampling, we explore an advanced similarity-based sampling strategy in Section~\ref{subs:bm25}.  An overall even higher standard deviation can be observed for MQL. We trace this additional variance to inconsistent output variations in LLMs (see also Section~\ref{sec:limitations}).

    \item \emph{LLMs have varying levels of knowledge across different query languages.} We suspect that this can be traced back to their training data. A large resource of such training data has been Stack Overflow~\cite{kabir2024stack}. When we search Stack Overflow for tags (indicated with []) related to our four query languages, we get the following numbers (15.08.2024): 
    [SQL]: 673K posts; [SPARQL]: 6K posts; [MongoDB, MQL]: 176K posts; [Cypher, Neo4J]: 33K posts.
    Relating the number of posts to our ``w/ schema 0-shot'' results (we want to leave the impact of few-shots ICL out of this), we see that SQL performs best (best model: 47.05\% ), Cypher and MQL perform average (best models: 34.45\% and 21.55\%), while SPARQL performs worst (best model: 3.3\%). These results correlate to the post frequency on Stack Overflow and support results by~\cite{kabir2024stack} that find a statistically significant impact on the correctness of LLM answers based on question popularity and recency. An exception to this pattern is MQL as it is under-performing Cypher. We suspect that this has to do with the fact that Cypher queries contain much fewer tokens and language keywords than MQL (only 25\% of tokens and 50\% of keywords, see Figure~\ref{fig:dataset_images}).

\end{itemize}

\begin{table}[ht] 
    \small
    \caption{Execution Accuracy of different LLMs \textbf{without} and \textbf{with schema information} for \textbf{test data}}  
    \label{tab:ea_test}  
    \resizebox{\linewidth}{!}{
    \setlength\tabcolsep{4pt}  
    \begin{tabular}{l c c c c c}  
    \toprule  
    \multirow{2}{*}{\textbf{Models}} & \multicolumn{2}{c}{\textbf{without schema}} & \multicolumn{3}{c}{\textbf{with schema}} \\   
    \cmidrule(lr){2-3} \cmidrule(lr){4-6}  
     & \textbf{w/o schema 1-shot} & \textbf{w/o schema 5-shot} & \textbf{w/ schema 0-shot} & \textbf{w/ schema 1-shot} & \textbf{w/ schema 5-shot} \\  
    \rowcolor{black!10!}\multicolumn{6}{c}{\textbf{\textit{SQL (PostgreSQL)}}} \\
    Llama3-8b
    & 4.20 \textbf{{\color{gray}\fontsize{7}{8.4}\selectfont(\textpm 5.6) }}
    & 10.81 \textbf{{\color{gray}\fontsize{7}{8.4}\selectfont(\textpm 9.89) }}
    & 22.55
    & 23.27 \textbf{{\color{gray}\fontsize{7}{8.4}\selectfont(\textpm 1.05)}}
    & 27.49 \textbf{{\color{gray}\fontsize{7}{8.4}\selectfont(\textpm 15.27)}} \\  
    Gemini 1.0 Pro
    & 4.47 \textbf{{\color{gray}\fontsize{7}{8.4}\selectfont(\textpm 4.88) }}
    & 21.65 \textbf{{\color{gray}\fontsize{7}{8.4}\selectfont(\textpm 11.10) }}
    & 38.60
    & 38.37 \textbf{{\color{gray}\fontsize{7}{8.4}\selectfont(\textpm 3.31)}}
    & 49.32 \textbf{{\color{gray}\fontsize{7}{8.4}\selectfont(\textpm 3.63)}} \\  
    GPT 3.5
    & 1.45 \textbf{{\color{gray}\fontsize{7}{8.4}\selectfont(\textpm 0.99) }}
    & 11.71 \textbf{{\color{gray}\fontsize{7}{8.4}\selectfont(\textpm 12.77) }}
    & 42.20
    & 48.92 \textbf{{\color{gray}\fontsize{7}{8.4}\selectfont(\textpm 6.72)}}
    & 56.30 \textbf{{\color{gray}\fontsize{7}{8.4}\selectfont(\textpm 2.36)}} \\
    Llama3-70b
    & 7.35 \textbf{{\color{gray}\fontsize{7}{8.4}\selectfont(\textpm 7.59) }}
    & 20.14 \textbf{{\color{gray}\fontsize{7}{8.4}\selectfont(\textpm 13.14) }}
    & 47.05
    & 51.06 \textbf{{\color{gray}\fontsize{7}{8.4}\selectfont(\textpm 1.75)}}
    & 57.50 \textbf{{\color{gray}\fontsize{7}{8.4}\selectfont(\textpm 2.91)}} \\
    \rowcolor{black!10!}\multicolumn{6}{c}{\textbf{\textit{SPARQL (GraphDB)}}} \\
    Llama3-8b
    & 3.09 \textbf{{\color{gray}\fontsize{7}{8.4}\selectfont(\textpm 2.70) }}
    & 4.18 \textbf{{\color{gray}\fontsize{7}{8.4}\selectfont(\textpm 9.04) }}
    & 0.05
    & 1.51 \textbf{{\color{gray}\fontsize{7}{8.4}\selectfont(\textpm 1.92)}}
    & 4.27  \textbf{{\color{gray}\fontsize{7}{8.4}\selectfont(\textpm 8.92)}} \\
    Gemini 1.0 Pro
    & 3.23 \textbf{{\color{gray}\fontsize{7}{8.4}\selectfont(\textpm 1.95) }}
    & 11.99 \textbf{{\color{gray}\fontsize{7}{8.4}\selectfont(\textpm 7.87) }}
    & 2.85
    & 7.76 \textbf{{\color{gray}\fontsize{7}{8.4}\selectfont(\textpm 4.65)}}
    & 26.32 \textbf{{\color{gray}\fontsize{7}{8.4}\selectfont(\textpm 5.60)}} \\
    GPT 3.5
    & 6.95 \textbf{{\color{gray}\fontsize{7}{8.4}\selectfont(\textpm 5.48) }}
    & 25.32 \textbf{{\color{gray}\fontsize{7}{8.4}\selectfont(\textpm 4.57) }}
    & 3.30
    & 7.88 \textbf{{\color{gray}\fontsize{7}{8.4}\selectfont(\textpm 4.78)}}
    & 23.58 \textbf{{\color{gray}\fontsize{7}{8.4}\selectfont(\textpm 8.09)}} \\
    Llama3-70b
    & 7.37 \textbf{{\color{gray}\fontsize{7}{8.4}\selectfont(\textpm 4.46) }}
    & 27.14 \textbf{{\color{gray}\fontsize{7}{8.4}\selectfont(\textpm 2.69) }}
    & 1.00
    & 10.26 \textbf{{\color{gray}\fontsize{7}{8.4}\selectfont(\textpm 6.89)}}
    & 30.49 \textbf{{\color{gray}\fontsize{7}{8.4}\selectfont(\textpm 1.82)}} \\
    \rowcolor{black!10!}\multicolumn{6}{c}{\textbf{\textit{Cypher (Neo4j)}}} \\
    Llama3-8b
    & 9.43 \textbf{{\color{gray}\fontsize{7}{8.4}\selectfont(\textpm 4.12) }}
    & 19.64 \textbf{{\color{gray}\fontsize{7}{8.4}\selectfont(\textpm 3.35) }}
    & 2.75
    & 15.31 \textbf{{\color{gray}\fontsize{7}{8.4}\selectfont(\textpm 11.28)}}
    & 34.89 \textbf{{\color{gray}\fontsize{7}{8.4}\selectfont(\textpm 5.34)}} \\
    Gemini 1.0 Pro
    & 13.80 \textbf{{\color{gray}\fontsize{7}{8.4}\selectfont(\textpm 1.67) }}
    & 22.91 \textbf{{\color{gray}\fontsize{7}{8.4}\selectfont(\textpm 1.38) }}
    & 23.45
    & 39.74 \textbf{{\color{gray}\fontsize{7}{8.4}\selectfont(\textpm 2.99)}}
    & 53.84 \textbf{{\color{gray}\fontsize{7}{8.4}\selectfont(\textpm 4.09)}} \\
    GPT 3.5
    & 10.37 \textbf{{\color{gray}\fontsize{7}{8.4}\selectfont(\textpm 4.84) }}
    & 18.08 \textbf{{\color{gray}\fontsize{7}{8.4}\selectfont(\textpm 1.05) }}
    & 16.35
    & 29.87 \textbf{{\color{gray}\fontsize{7}{8.4}\selectfont(\textpm 3.44)}}
    & 41.12 \textbf{{\color{gray}\fontsize{7}{8.4}\selectfont(\textpm 2.85)}} \\
    Llama3-70b
    & 16.04 \textbf{{\color{gray}\fontsize{7}{8.4}\selectfont(\textpm 2.40) }}
    & 25.25 \textbf{{\color{gray}\fontsize{7}{8.4}\selectfont(\textpm 5.10) }}
    & 34.45
     & 43.06 \textbf{{\color{gray}\fontsize{7}{8.4}\selectfont(\textpm 4.53)}}
     & 57.07 \textbf{{\color{gray}\fontsize{7}{8.4}\selectfont(\textpm 4.41)}} \\
    \rowcolor{black!10!}\multicolumn{6}{c}{\textbf{\textit{MQL (MongoDB)}}} \\
    Llama3-8b
    & 2.64 \textbf{{\color{gray}\fontsize{7}{8.4}\selectfont(\textpm 3.35) }}
    & 4.62 \textbf{{\color{gray}\fontsize{7}{8.4}\selectfont(\textpm 6.56) }}
    & 9.45
    & 6.71 \textbf{{\color{gray}\fontsize{7}{8.4}\selectfont(\textpm 6.55)}}
    & 11.33 \textbf{{\color{gray}\fontsize{7}{8.4}\selectfont(\textpm 15.06)}} \\  
    Gemini 1.0 Pro
    & 5.25 \textbf{{\color{gray}\fontsize{7}{8.4}\selectfont(\textpm 2.47) }}
    & 13.25 \textbf{{\color{gray}\fontsize{7}{8.4}\selectfont(\textpm 3.25) }}
    & 3.40
    & 18.53 \textbf{{\color{gray}\fontsize{7}{8.4}\selectfont(\textpm 1.67)}}
    & 30.65 \textbf{{\color{gray}\fontsize{7}{8.4}\selectfont(\textpm 7.19)}} \\  
    GPT 3.5
    & 1.49 \textbf{{\color{gray}\fontsize{7}{8.4}\selectfont(\textpm 3.30) }}
    & 5.36 \textbf{{\color{gray}\fontsize{7}{8.4}\selectfont(\textpm 5.17) }}
    & 3.50
    & 26.26 \textbf{{\color{gray}\fontsize{7}{8.4}\selectfont(\textpm 13.64)}}
    & 35.06 \textbf{{\color{gray}\fontsize{7}{8.4}\selectfont(\textpm 15.74)}} \\ 
    Llama3-70b
    & 8.86 \textbf{{\color{gray}\fontsize{7}{8.4}\selectfont(\textpm 2.09) }}
    & 17.91 \textbf{{\color{gray}\fontsize{7}{8.4}\selectfont(\textpm 4.52) }}
    & 21.55
    & 33.83 \textbf{{\color{gray}\fontsize{7}{8.4}\selectfont(\textpm 8.54)}}
    & 40.35 \textbf{{\color{gray}\fontsize{7}{8.4}\selectfont(\textpm 17.03)}} \\ 
    \bottomrule  
    \end{tabular}}
        \rmspace
\end{table}

\subsection{Per-category Results}

\begin{figure}[ht]
    \centering
    \includegraphics[width=1.0\textwidth]{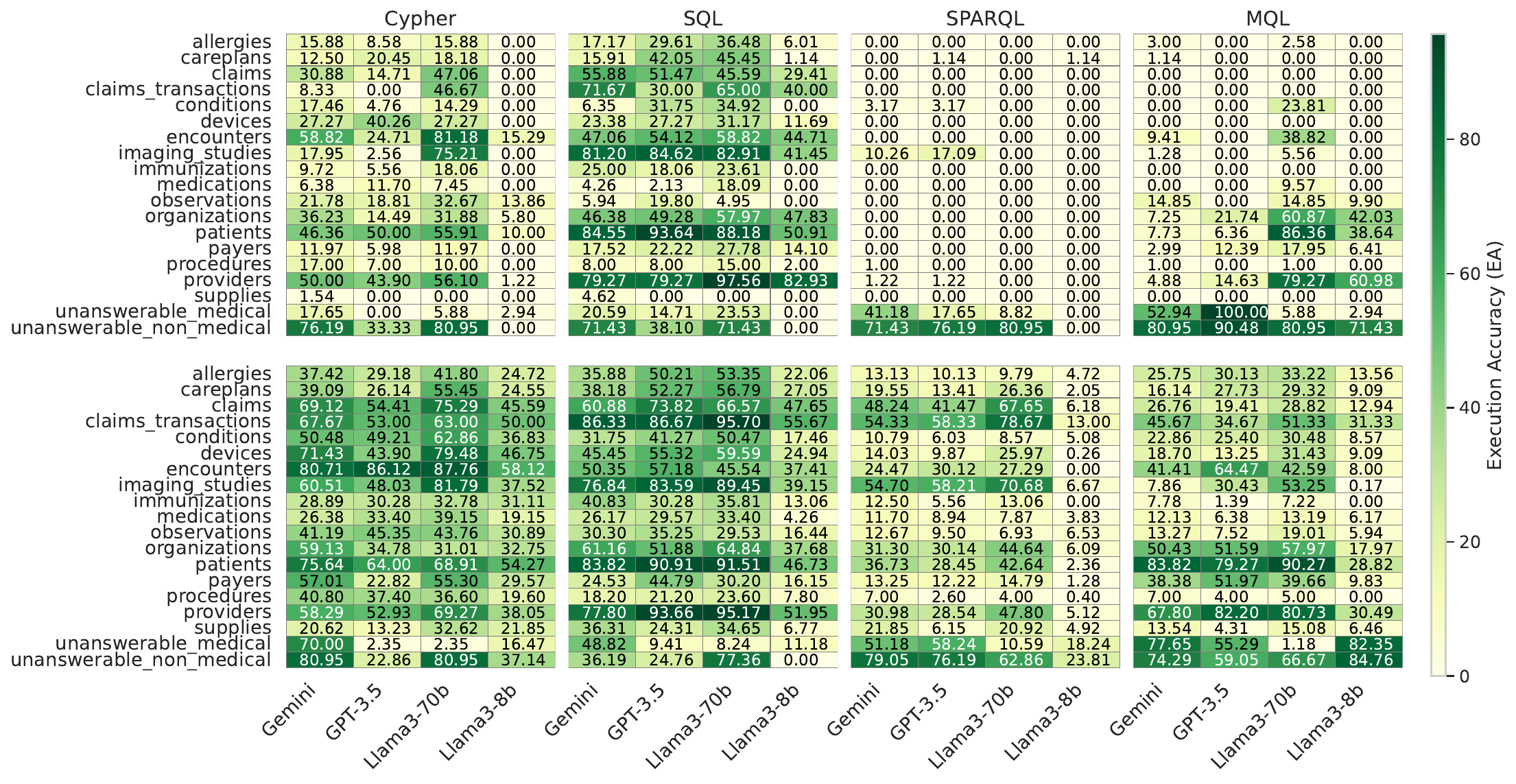}
    \caption{Execution Accuracy (EA) for our 19 different question categories for \textit{w/ schema 0-shot} (top) and \textit{w/ schema 5-shot} (bottom).}
    \label{fig:accuracy_category}
    \rmspace
\end{figure}

Next, we analyze the performance on a per-category level based on the entity-tagged template questions. For that, we look at the results for \textit{w/ schema 0-shot} and \textit{w/ schema 5-shot} to observe the impact of ICL through few-shot examples for our 19 question categories, our four query languages, and four LLMs.
Figure~\ref{fig:accuracy_category} (top) shows the execution accuracy based on the w/ schema 0-shot results, while Figure~\ref{fig:accuracy_category} (bottom) shows the mean results for w/ schema 5-shot experiments.

We observe a clear difference between high and low-resource query languages. Despite the available schema, correct SPARQL and MQL queries can mostly not be generated for all LLMs without few-shot examples (top, 3rd, and 4th columns). For these low-resource languages, performance improves substantially with ICL. We also observe differences that can be traced to model size and potential training data.
Llama3-8b, the smallest model, struggles even with examples to produce correct SPARQL queries. Both Llama models seem to encode more knowledge about MQL than the other LLMs (highest schema 0-shot results).
For MQL, we see the highest performance for questions related to top-level entities (i.e., not nested objects), such as \textit{patients}, \textit{organizations}, \textit{providers}, \textit{payers}.
For non-answerable questions, we observe that non-medical questions have a higher accuracy than medical ones. The capability of an LLM to recognize unanswerable questions varies across query languages, even with the same prompt instruction.

\subsection{Similarity-based few-shot sample selection}
\label{subs:bm25}
Last, we implement a similarity-based approach using a BM25~\cite{robertson2009probabilistic} retriever that, on a per-question basis, retrieves the five most similar question-query pairs from the training data. We use these question pairs as few-shot examples instead of the fixed few-shots used before. We use the w/ schema prompt with GPT-3.5. This greatly improves results to up to 88.55\% for SQL (see Table \ref{tab:icl_exp}), which is in line with related results for the EHRSQL dataset in EHRXQA~\cite{bae2024ehrxqa}.
As an end-to-end comparison, EHRXQA achieves an execution accuracy of 92.9\%, an improvement by 62.9 points from their fixed-shot strategy. This is consistent with our results, where SQL execution accuracy improves by 32.25 points to 88.55\%. However, we can also see that even with a state-of-the-art ICL method, SPARQL, Cypher, and MQL cannot reach the same performance as SQL (MQL has the highest score with 78.70\%). This indicates the need for more research, such as developing better ICL methods and potentially schema encodings. Overall, it reinforces the motivation and strengthens the necessity of our work in terms of creating a new multi-model Text-to-Query benchmark to work on these problems and to extend the research scope from ``Text-to-SQL'' to multi-model ``Text-to-Query''.

\begin{table}[hb]
\caption{
Advanced In-Context Learning (ICL) Method: BM25-based few-shot selection (5-shot) determined by question similarity to the training data.
}
\label{tab:icl_exp}
\centering
\resizebox{0.6\hsize}{!}{
\begin{tabular}{lrr}
\toprule
\textbf{Query Language} & \textbf{EA in \%}  & \textbf{Improvement to fixed 5-shot}\\
\midrule
SQL (PostgreSQL) & 88.55 & +32.25\\
SPARQL (GraphDB) & 75.75 & +52.17 \\
Cypher (Neo4j) & 78.30 & +37.18\\
MQL (MongoDB) & 78.70 & +43.64\\
\bottomrule
\end{tabular}}
\end{table}

\section{Discussion and Limitations}
\label{sec:limitations}

While SM3-Text-to-Query is the first benchmark across four different query languages, we note several limitations, some of which provide the potential for further research. 

First, compared to other datasets~\cite{lee2022ehrsql}, our question templates were only created with guidance from health professionals, but not directly formulated by them. Our dataset is synthetic, based on simulated patient data, with the benefit of flexibility and no privacy issues. In the future, we plan to extend our question set through crowd-sourcing as part of an ongoing Swiss digital health project~\cite{dizh}. 

Second, SM3 currently only contains English questions and database values. We believe that the addition of multilingual questions and databases could be a valuable 
extension to our benchmark.

Third, while our questions cover all main entities in the dataset, their corresponding queries might be too easy in some query languages (e.g., SQL and Cypher require, on average, less than 10 tokens; see Figure~\ref{fig:dataset_images}). There is the potential to include temporal templates as in~\cite{lee2022ehrsql} to increase query complexity.

Last, we experience that LLMs exhibit large output variations for the same prompt and their generations can be inconsistent across query languages. We implemented extensive data-cleaning logic to extract the predicted query from the LLM output. These outputs varied across models but also across languages: while GPT-3.5 follows instructions well for SQL, it does not for MQL, despite the same structured prompt (see Appendix~\ref{app:sec:issues} for examples of encountered issues).
Further research should focus on prompt optimization across database models, potentially using LLMs as optimizers~\cite{yang2023large}.

\textbf{Potential negative societal impacts.} There are no direct negative societal impacts associated with our dataset. It will help researchers and people in industry to improve their Text-to-Query systems, thereby democratizing data access to wider user groups.

\section{Conclusion}

This paper provides, to the best of our knowledge, the first multi-model Text-to-Query dataset and benchmark that allows for the evaluation of Text-to-Query systems across three core database models (relational, graph, document) and four query languages (SQL, SPARQL, Cypher, MQL). Our dataset is based on synthetic medical data generated through Synthea~\cite{synthea_international}, follows an international medical ontology standard (SNOMED~\cite{SNOMED-International:2024aa}), and can be easily extended through further template questions or by exchanging the synthetic data through standard-conform real patient data. SM3-Text-to-Query will be essential to develop and test the next generation of Text-to-Query systems that appear with increasing frequency thanks to the progress in transformer-based large language models. All our code and data are available at \url{https://github.com/jf87/SM3-Text-to-Query}.

\acksection
We thank all constructive comments from the anonymous reviewers.
This work has been supported by OpenAI’s Researcher Access Program.

\bibliographystyle{abbrvnat}
{
\small
\bibliography{bibliography}
}

\section*{Checklist}


\begin{enumerate}

\item For all authors...
\begin{enumerate}
  \item Do the main claims made in the abstract and introduction accurately reflect the paper's contributions and scope?
    \answerYes{}
  \item Did you describe the limitations of your work?
    \answerYes{See Section~\ref{sec:limitations}.}
  \item Did you discuss any potential negative societal impacts of your work?
    \answerYes{See Section~\ref{sec:limitations}.}
  \item Have you read the ethics review guidelines and ensured that your paper conforms to them?
    \answerYes{Our research does not involve human participants. Our dataset is purely synthetic and does not contain personal data. Synthea, the patient data simulator is published under an open-source Apache License.}
\end{enumerate}

\item If you are including theoretical results...
\begin{enumerate}
  \item Did you state the full set of assumptions of all theoretical results?
    \answerNA{Benchmark and dataset paper.}
	\item Did you include complete proofs of all theoretical results?
    \answerNA{Benchmark and dataset paper.}
\end{enumerate}

\item If you ran experiments (e.g. for benchmarks)...
\begin{enumerate}
  \item Did you include the code, data, and instructions needed to reproduce the main experimental results (either in the supplemental material or as a URL)?
    \answerYes{All our code and data is available in the supplemental material and will be available to the general public after acceptance of the paper. Our goal is to have a public and reproducible benchmark that is used by as many people as possible.}
  \item Did you specify all the training details (e.g., data splits, hyperparameters, how they were chosen)?
     \answerYes{Yes, see descriptions in Section~\ref{sec:dataset} and Section~\ref{sub:experimental-setup}}
	\item Did you report error bars (e.g., with respect to the random seed after running experiments multiple times)?
    \answerYes{All our results for experiments with multiple runs (e.g., few-shot experiments) contain the standard deviation indicated with \textpm.}
	\item Did you include the total amount of compute and the type of resources used (e.g., type of GPUs, internal cluster, or cloud provider)?
     \answerYes{Yes, see description in Section~\ref{sub:experimental-setup} and about the setup for the PostgreSQL, GraphDB, Neo4j and GraphDB databases Appendix~\ref{app:efficiency}.}

\end{enumerate}

\item If you are using existing assets (e.g., code, data, models) or curating/releasing new assets...
\begin{enumerate}
  \item If your work uses existing assets, did you cite the creators?
    \answerYes{For example Synthea~\cite{synthea_github} and SNOMED~\cite{SNOMED-International:2024aa}.}
  \item Did you mention the license of the assets?
    \answerYes{Our dataset and code will be released under CC-BY-SA license. For further details, see answerYes.}
  \item Did you include any new assets either in the supplemental material or as a URL?
    \answerYes{Yes, see supplemental material and link to Git repository.}
  \item Did you discuss whether and how consent was obtained from people whose data you're using/curating?
    \answerNA{We have generated our own synthetic data with the Synthea open-source asset. It is not data obtained from humans (e.g., personal data).}
  \item Did you discuss whether the data you are using/curating contains personally identifiable information or offensive content?
    \answerNA{See answer before, our data is purely synthetic. The annotated question/query pairs do not contain personal data.}
\end{enumerate}

\item If you used crowdsourcing or conducted research with human subjects...
\begin{enumerate}
  \item Did you include the full text of instructions given to participants and screenshots, if applicable?
    \answerNA{No crowdsourcing or research with human subjects was done.}
  \item Did you describe any potential participant risks, with links to Institutional Review Board (IRB) approvals, if applicable?
    \answerNA{No crowdsourcing or research with human subjects was done.}
  \item Did you include the estimated hourly wage paid to participants and the total amount spent on participant compensation?
    \answerNA{No crowdsourcing or research with human subjects was done.}
\end{enumerate}

\end{enumerate}


\clearpage
\appendix

\section{Appendix}

We include below technical appendices (such as the evaluation results on the development dataset and efficiency results). The additional material that supports our dataset and benchmark documentation is provided in the supplement material.

\subsection{Text-to-Query Accuracy Development Data}

Table~\ref{tab:ea_dev} summarizes the Execution Accuracy (EA) results of the different models, prompts, and query languages for the \textit{development} dataset - equivalent to Table~\ref{tab:ea_test} for the \textit{test} data in the main part of the paper. However, due to timing and cost constraints, we have only performed three runs and used a smaller dataset for Llama3-70b. The results from the development set show similar patterns and insights as we have described in Section~\ref{sec:evaluation}.

\begin{table} [h]
    \small
    \caption{Execution Accuracy of different LLMs \textbf{without} and \textbf{with schema information} for \textbf{dev data}.} 
    \label{tab:ea_dev}  
    \resizebox{\linewidth}{!}{
    \setlength\tabcolsep{4pt}  
    \begin{tabular}{l c c c c c}  
    \toprule  
    \multirow{2}{*}{\textbf{Models}} & \multicolumn{2}{c}{\textbf{without schema}} & \multicolumn{3}{c}{\textbf{with schema}} \\   
    \cmidrule(lr){2-3} \cmidrule(lr){4-6}  
     & \textbf{w/o schema 1-shot} & \textbf{w/o schema 5-shot} & \textbf{w/ schema 0-shot} & \textbf{w/ schema 1-shot} & \textbf{w/ schema 5-shot} \\  
    \rowcolor{black!10!}\multicolumn{6}{c}{\textbf{\textit{SQL (PostgreSQL)}}} \\
    Llama3-8b
     & 9.37 \textbf{{\color{gray}\fontsize{7}{8.4}\selectfont(\textpm1.64)}}
     & 16.07 \textbf{{\color{gray}\fontsize{7}{8.4}\selectfont(\textpm 3.67)}}
     & 21.50
     & 23.82 \textbf{{\color{gray}\fontsize{7}{8.4}\selectfont(\textpm 0.42) }}
     & 34.48 \textbf{{\color{gray}\fontsize{7}{8.4}\selectfont(\textpm 4.46)}}\\  
    Gemini 1.0 Pro
     & 7.70 \textbf{{\color{gray}\fontsize{7}{8.4}\selectfont(\textpm 4.21)}} 
     & 12.97 \textbf{{\color{gray}\fontsize{7}{8.4}\selectfont(\textpm 11.27)}} 
     & 37.90
     & 30.38 \textbf{{\color{gray}\fontsize{7}{8.4}\selectfont(\textpm 15.29)}}
     & 50.38 \textbf{{\color{gray}\fontsize{7}{8.4}\selectfont(\textpm 4.45)}} \\  
    GPT 3.5
    & 8.88 \textbf{{\color{gray}\fontsize{7}{8.4}\selectfont(\textpm 1.76)}} 
    & 18.43 \textbf{{\color{gray}\fontsize{7}{8.4}\selectfont(\textpm  6.23)}} 
    & 41.10
    & 52.40 \textbf{{\color{gray}\fontsize{7}{8.4}\selectfont(\textpm 5.53)}}
    & 56.83 \textbf{{\color{gray}\fontsize{7}{8.4}\selectfont(\textpm 1.38)}} \\
    Llama3-70b\footnote{Llama3-70b were only tested with 200 samples.}
     & 11.83 \textbf{{\color{gray}\fontsize{7}{8.4}\selectfont(\textpm 6.25)}} 
     & 27.00 \textbf{{\color{gray}\fontsize{7}{8.4}\selectfont(\textpm 8.85)}}
     & 45.50
     & 26.67 \textbf{{\color{gray}\fontsize{7}{8.4}\selectfont(\textpm 3.53)}}
     & 42.17 \textbf{{\color{gray}\fontsize{7}{8.4}\selectfont(\textpm 0.94)}} \\
    \rowcolor{black!10!}\multicolumn{6}{c}{\textbf{\textit{SPARQL (GraphDB)}}} \\
    Llama3-8b
    & 4.53 \textbf{{\color{gray}\fontsize{7}{8.4}\selectfont(\textpm 3.49)}} 
    & 11.37 \textbf{{\color{gray}\fontsize{7}{8.4}\selectfont(\textpm 10.10)}} 
    & 0.00
    & 5.17 \textbf{{\color{gray}\fontsize{7}{8.4}\selectfont(\textpm 3.02)}}
    & 18.12 \textbf{{\color{gray}\fontsize{7}{8.4}\selectfont(\textpm 2.57 )}}\\
    Gemini 1.0 Pro
    & 4.68 \textbf{{\color{gray}\fontsize{7}{8.4}\selectfont(\textpm 2.26)}} 
    & 19.37 \textbf{{\color{gray}\fontsize{7}{8.4}\selectfont(\textpm 3.09)}} 
    & 2.90
    & 8.52 \textbf{{\color{gray}\fontsize{7}{8.4}\selectfont(\textpm 6.75)}}
    & 22.28 \textbf{{\color{gray}\fontsize{7}{8.4}\selectfont(\textpm12.78)}}\\
    GPT 3.5
    & 9.50 \textbf{{\color{gray}\fontsize{7}{8.4}\selectfont(\textpm 6.87)}} 
    & 27.90 \textbf{{\color{gray}\fontsize{7}{8.4}\selectfont(\textpm 5.24)}}
    & 3.80
    & 10.48 \textbf{{\color{gray}\fontsize{7}{8.4}\selectfont(\textpm 5.61)}}
    & 27.97 \textbf{{\color{gray}\fontsize{7}{8.4}\selectfont(\textpm 4.58)}} \\
    Llama3-70b\footnotemark[\value{footnote}]
    & 5.00 \textbf{{\color{gray}\fontsize{7}{8.4}\selectfont(\textpm 6.61)}} 
    & 22.67 \textbf{{\color{gray}\fontsize{7}{8.4}\selectfont(\textpm 5.11)}} 
    & 0.00
    & 9.00 \textbf{{\color{gray}\fontsize{7}{8.4}\selectfont(\textpm 7.57)}}
    & 25.17 \textbf{{\color{gray}\fontsize{7}{8.4}\selectfont(\textpm 4.01)}}\\
    \rowcolor{black!10!}\multicolumn{6}{c}{\textbf{\textit{Cypher (Neo4j)}}} \\
    Llama3-8b
    & 9.22 \textbf{{\color{gray}\fontsize{7}{8.4}\selectfont(\textpm 4.71)}} 
    & 17.07 \textbf{{\color{gray}\fontsize{7}{8.4}\selectfont(\textpm 3.87)}}
    & 1.65
    & 20.88 \textbf{{\color{gray}\fontsize{7}{8.4}\selectfont(\textpm 5.39)}}
    & 37.33 \textbf{{\color{gray}\fontsize{7}{8.4}\selectfont(\textpm 3.69)}}\\
    Gemini 1.0 Pro
    & 11.83 \textbf{{\color{gray}\fontsize{7}{8.4}\selectfont(\textpm 2.26)}} 
    & 23.23 \textbf{{\color{gray}\fontsize{7}{8.4}\selectfont(\textpm 1.16)}} 
    & 22.45
    & 38.85 \textbf{{\color{gray}\fontsize{7}{8.4}\selectfont(\textpm 3.13)}}
    & 49.73 \textbf{{\color{gray}\fontsize{7}{8.4}\selectfont(\textpm 2.45 )}} \\
    GPT 3.5
    & 7.95 \textbf{{\color{gray}\fontsize{7}{8.4}\selectfont(\textpm 3.99)}} 
    & 18.52 \textbf{{\color{gray}\fontsize{7}{8.4}\selectfont(\textpm 1.53)}} 
    & 16.35
    & 29.97 \textbf{{\color{gray}\fontsize{7}{8.4}\selectfont(\textpm 4.82)}}
    & 39.70 \textbf{{\color{gray}\fontsize{7}{8.4}\selectfont(\textpm 1.78)}} \\
    Llama3-70b\footnotemark[\value{footnote}]
    & 13.83 \textbf{{\color{gray}\fontsize{7}{8.4}\selectfont(\textpm 3.01)}} 
    & 18.67 \textbf{{\color{gray}\fontsize{7}{8.4}\selectfont(\textpm  3.55)}} 
    & 29.00
    & 22.50 \textbf{{\color{gray}\fontsize{7}{8.4}\selectfont(\textpm 19.11)}}
    & 45.33 \textbf{{\color{gray}\fontsize{7}{8.4}\selectfont(\textpm 5.01)}} \\
    \rowcolor{black!10!}\multicolumn{6}{c}{\textbf{\textit{MQL (MongoDB)}}} \\
    Llama3-8b
    & 4.73 \textbf{{\color{gray}\fontsize{7}{8.4}\selectfont(\textpm 2.91) }}
    & 12.10 \textbf{{\color{gray}\fontsize{7}{8.4}\selectfont(\textpm 2.80) }}
    & 10.25
    & 13.18 \textbf{{\color{gray}\fontsize{7}{8.4}\selectfont(\textpm 0.95)}}
    & 26.47 \textbf{{\color{gray}\fontsize{7}{8.4}\selectfont(\textpm 5.03)}} \\  
    Gemini 1.0 Pro
    & 7.10 \textbf{{\color{gray}\fontsize{7}{8.4}\selectfont(\textpm 0.65) }}
    & 15.65 \textbf{{\color{gray}\fontsize{7}{8.4}\selectfont(\textpm 4.53) }}
    & 5.95
    & 21.60 \textbf{{\color{gray}\fontsize{7}{8.4}\selectfont(\textpm 1.72)}}
    & 37.97 \textbf{{\color{gray}\fontsize{7}{8.4}\selectfont(\textpm 2.28)}} \\  
    GPT 3.5
    & 3.15 \textbf{{\color{gray}\fontsize{7}{8.4}\selectfont(\textpm 0.26) }}
    & 4.88 \textbf{{\color{gray}\fontsize{7}{8.4}\selectfont(\textpm 1.89) }}
    & 9.45
    & 24.93 \textbf{{\color{gray}\fontsize{7}{8.4}\selectfont(\textpm 14.28)}}
    & 33.08 \textbf{{\color{gray}\fontsize{7}{8.4}\selectfont(\textpm 17.30)}} \\ 
    Llama3-70b\footnotemark[\value{footnote}]
    & 8.17 \textbf{{\color{gray}\fontsize{7}{8.4}\selectfont(\textpm 1.15) }}
    & 19.00 \textbf{{\color{gray}\fontsize{7}{8.4}\selectfont(\textpm 1.73) }}
    & 22.50
    & 23.67 \textbf{{\color{gray}\fontsize{7}{8.4}\selectfont(\textpm 11.86)}}
    & 42.50 \textbf{{\color{gray}\fontsize{7}{8.4}\selectfont(\textpm 3.28)}} \\ 
    \bottomrule  
    \end{tabular}}
    \parbox{\linewidth}{\scriptsize \textit{\footnotemark[\value{footnote}] Llama3-70b was only tested with 200 test and 200 dev random samples from a uniform distribution due to cost/runtime constraints.}}
\end{table}

\clearpage

\subsection{Text-to-Query Efficiency}
\label{app:efficiency}

The run-time efficiency of queries against single database systems, a core problem of query optimization, has recently also drawn attention from the text-to-SQL community. 
An interesting aspect is to compare the runtime of the manually created ground truth queries with the generated queries.

Here, we show results for the recently proposed Valid Efficiency Score (VES)~\cite{li2024can}, which measures Text-to-Query Efficiency.
We execute all queries three times. The respective databases are running on virtual machines on an OpenStack cluster with the same specs (8 cores 16GB RAM). We omit the LLM inference time, as this time varies substantively by parameters that we cannot fully control (e.g., current OpenAI, Google, Groq server load). Moreover, our focus is on query execution time within a database system, which is independent of the machine learning inference time for generating the queries.

\textbf{Valid Efficiency Score (VES).}
Pioneered by \textsc{Bird}~\cite{li2024can}, the VES score aims to include the efficiency of the generated query together with the Execution Accuracy (EA). 

\begin{equation}
\small
\mathrm{VES}=\frac{\Sigma_{n=1}^N \mathbbm{1}(r_n, \hat{r}_n)\cdot\mathbf{R}(Y_{n}, \hat{Y}_{n})}{N}, \quad \mathbf{R}(Y_{n}, \hat{Y}_{n}) = \sqrt{\frac{\mathbf{E}(Y_n)}{\mathbf{E}(\hat{Y}_n)}}
\label{eq4}
\end{equation}
where $\mathbf{R(\cdot)}$ denotes the relative execution efficiency of the predicted query in comparison to the ground-truth query. $\mathbf{E}(\cdot)$ is a function to measure the absolute execution efficiency for each query in a given environment, e.g. execution time in milliseconds. For further details, we refer the reader to~\cite{li2024can}.
Table~\ref{tab:ves-test} and Table~\ref{tab:ves-dev} depict the VES results for the test and dev dataset, respectively. 

\begin{table}[h] 
    \small
    \caption{Valid Efficiency Score (VES) of different LLMs \textbf{without} and \textbf{with schema information} for \textbf{test data}.}  
    \label{tab:ves-test}  
    \resizebox{\linewidth}{!}{
    \setlength\tabcolsep{4pt}  
    \begin{tabular}{l c c c c c}  
    \toprule  
    \multirow{2}{*}{\textbf{Models}} & \multicolumn{2}{c}{\textbf{without schema}} & \multicolumn{3}{c}{\textbf{with schema}} \\   
    \cmidrule(lr){2-3} \cmidrule(lr){4-6}  
     & \textbf{w/o schema 1-shot} & \textbf{w/o schema 5-shot} & \textbf{w/ schema 0-shot} & \textbf{w/ schema 1-shot} & \textbf{w/ schema 5-shot} \\  
    \rowcolor{black!10!}\multicolumn{6}{c}{\textbf{\textit{SQL (PostgreSQL)}}} \\
    Llama3-8b
    & 0.85 \textbf{{\color{gray}\fontsize{7}{8.4}\selectfont(\textpm 1.18) }}
    & 2.64 \textbf{{\color{gray}\fontsize{7}{8.4}\selectfont(\textpm  2.52) }}
    & 5.57
    & 5.77 \textbf{{\color{gray}\fontsize{7}{8.4}\selectfont(\textpm 0.74)}}
    & 7.35 \textbf{{\color{gray}\fontsize{7}{8.4}\selectfont(\textpm 4.41)}} \\ 
    Gemini 1.0 Pro
    & 0.77 \textbf{{\color{gray}\fontsize{7}{8.4}\selectfont(\textpm 1.07) }}
    & 5.28 \textbf{{\color{gray}\fontsize{7}{8.4}\selectfont(\textpm 3.19) }}
    & 10.94
    & 10.15 \textbf{{\color{gray}\fontsize{7}{8.4}\selectfont(\textpm 1.75)}}
    & 13.38 \textbf{{\color{gray}\fontsize{7}{8.4}\selectfont(\textpm 1.63)}} \\  
    GPT 3.5
    & 0.00 \textbf{{\color{gray}\fontsize{7}{8.4}\selectfont(\textpm 0.00) }}
    & 2.33 \textbf{{\color{gray}\fontsize{7}{8.4}\selectfont(\textpm 3.27) }}
    & 10.81
    & 12.94 \textbf{{\color{gray}\fontsize{7}{8.4}\selectfont(\textpm 2.20)}}
    & 16.65 \textbf{{\color{gray}\fontsize{7}{8.4}\selectfont(\textpm 0.94)}} \\
    Llama3-70b 
    & 1.81 \textbf{{\color{gray}\fontsize{7}{8.4}\selectfont(\textpm 2.03) }}
    & 5.71 \textbf{{\color{gray}\fontsize{7}{8.4}\selectfont(\textpm 3.83) }}
    & 12.85
    & 13.98 \textbf{{\color{gray}\fontsize{7}{8.4}\selectfont(\textpm 0.73)}}
    & 16.64 \textbf{{\color{gray}\fontsize{7}{8.4}\selectfont(\textpm 1.61)}} \\
    \rowcolor{black!10!}\multicolumn{6}{c}{\textbf{\textit{SPARQL (GraphDB)}}} \\
    Llama3-8b
    & 2.89 \textbf{{\color{gray}\fontsize{7}{8.4}\selectfont(\textpm 2.51) }}
    & 3.58 \textbf{{\color{gray}\fontsize{7}{8.4}\selectfont(\textpm 7.95) }}
    & 0.03
    & 1.47 \textbf{{\color{gray}\fontsize{7}{8.4}\selectfont(\textpm  1.88)}}
    & 3.53  \textbf{{\color{gray}\fontsize{7}{8.4}\selectfont(\textpm 7.85)}} \\
    Gemini 1.0 Pro
    & 1.28 \textbf{{\color{gray}\fontsize{7}{8.4}\selectfont(\textpm 1.22) }}
    & 9.93 \textbf{{\color{gray}\fontsize{7}{8.4}\selectfont(\textpm 7.46) }}
    & 1.29
    & 6.26 \textbf{{\color{gray}\fontsize{7}{8.4}\selectfont(\textpm  4.70)}}
    & 23.00 \textbf{{\color{gray}\fontsize{7}{8.4}\selectfont(\textpm 4.78)}} \\
    GPT 3.5
    & 4.50 \textbf{{\color{gray}\fontsize{7}{8.4}\selectfont(\textpm 5.00) }}
    & 22.18 \textbf{{\color{gray}\fontsize{7}{8.4}\selectfont(\textpm 5.19 	) }}
    & 2.17
    & 6.05 \textbf{{\color{gray}\fontsize{7}{8.4}\selectfont(\textpm 4.64)}}
    & 20.94 \textbf{{\color{gray}\fontsize{7}{8.4}\selectfont(\textpm 7.89)}} \\
    Llama3-70b 
    & 6.96 \textbf{{\color{gray}\fontsize{7}{8.4}\selectfont(\textpm 4.42) }}
    & 24.89 \textbf{{\color{gray}\fontsize{7}{8.4}\selectfont(\textpm 2.09 	) }}
    & 0.00
    & 9.14 \textbf{{\color{gray}\fontsize{7}{8.4}\selectfont(\textpm 6.44)}}
    & 27.81 \textbf{{\color{gray}\fontsize{7}{8.4}\selectfont(\textpm 1.95)}} \\
    \rowcolor{black!10!}\multicolumn{6}{c}{\textbf{\textit{Cypher (Neo4j)}}} \\
    Llama3-8b
    & 7.44 \textbf{{\color{gray}\fontsize{7}{8.4}\selectfont(\textpm 2.29) }}
    & 15.32 \textbf{{\color{gray}\fontsize{7}{8.4}\selectfont(\textpm 2.43) }}
    & 2.36
    & 12.53 \textbf{{\color{gray}\fontsize{7}{8.4}\selectfont(\textpm 10.59)}}
    & 26.92 \textbf{{\color{gray}\fontsize{7}{8.4}\selectfont(\textpm 8.03)}} \\
    Gemini 1.0 Pro
    & 10.70 \textbf{{\color{gray}\fontsize{7}{8.4}\selectfont(\textpm 2.88) }}
    & 17.51 \textbf{{\color{gray}\fontsize{7}{8.4}\selectfont(\textpm 3.25) }}
    & 19.43
    & 30.25 \textbf{{\color{gray}\fontsize{7}{8.4}\selectfont(\textpm 7.23)}}
    & 40.52 \textbf{{\color{gray}\fontsize{7}{8.4}\selectfont(\textpm 6.16)}} \\
    GPT 3.5
    & 6.33 \textbf{{\color{gray}\fontsize{7}{8.4}\selectfont(\textpm 4.01) }}
    & 12.92 \textbf{{\color{gray}\fontsize{7}{8.4}\selectfont(\textpm 2.59) }}
    & 13.28
    & 22.26 \textbf{{\color{gray}\fontsize{7}{8.4}\selectfont(\textpm 6.78)}}
    & 29.48 \textbf{{\color{gray}\fontsize{7}{8.4}\selectfont(\textpm 4.46)}} \\
    Llama3-70b 
    & 11.38 \textbf{{\color{gray}\fontsize{7}{8.4}\selectfont(\textpm 1.22) }}
    & 17.55 \textbf{{\color{gray}\fontsize{7}{8.4}\selectfont(\textpm 2.74) }}
    & 23.57
    & 29.45 \textbf{{\color{gray}\fontsize{7}{8.4}\selectfont(\textpm 3.31)}}
    & 39.90 \textbf{{\color{gray}\fontsize{7}{8.4}\selectfont(\textpm 2.44)}} \\
    \rowcolor{black!10!}\multicolumn{6}{c}{\textbf{\textit{MQL (MongoDB)}}} \\
    Llama3-8b
    & 2.98 \textbf{{\color{gray}\fontsize{7}{8.4}\selectfont(\textpm 3.78) }}
    & 5.18 \textbf{{\color{gray}\fontsize{7}{8.4}\selectfont(\textpm 7.31) }}
    & 10.54
    & 7.54 \textbf{{\color{gray}\fontsize{7}{8.4}\selectfont(\textpm 7.33)}}
    & 12.66 \textbf{{\color{gray}\fontsize{7}{8.4}\selectfont(\textpm 16.84)}} \\  
    Gemini 1.0 Pro
    & 5.97 \textbf{{\color{gray}\fontsize{7}{8.4}\selectfont(\textpm 2.81) }}
    & 15.04 \textbf{{\color{gray}\fontsize{7}{8.4}\selectfont(\textpm 3.68) }}
    & 3.78
    & 20.88 \textbf{{\color{gray}\fontsize{7}{8.4}\selectfont(\textpm 1.70)}}
    & 34.07 \textbf{{\color{gray}\fontsize{7}{8.4}\selectfont(\textpm 7.39)}} \\  
    GPT 3.5
    & 1.71 \textbf{{\color{gray}\fontsize{7}{8.4}\selectfont(\textpm 3.79) }}
    & 6.05 \textbf{{\color{gray}\fontsize{7}{8.4}\selectfont(\textpm 5.88) }}
    & 3.92
    & 29.04 \textbf{{\color{gray}\fontsize{7}{8.4}\selectfont(\textpm 14.88)}}
    & 38.90 \textbf{{\color{gray}\fontsize{7}{8.4}\selectfont(\textpm 17.47)}} \\ 
    Llama3-70b
    & 10.10 \textbf{{\color{gray}\fontsize{7}{8.4}\selectfont(\textpm 2.40) }}
    & 20.36 \textbf{{\color{gray}\fontsize{7}{8.4}\selectfont(\textpm 5.02) }}
    & 23.65
    & 37.40 \textbf{{\color{gray}\fontsize{7}{8.4}\selectfont(\textpm 8.91)}}
    & 45.48 \textbf{{\color{gray}\fontsize{7}{8.4}\selectfont(\textpm 18.95)}} \\ 
    \bottomrule  
    \end{tabular}}
\end{table}

\begin{table}[h]
    \small
    \caption{Valid Efficiency Score (VES) of different LLMs \textbf{without} and \textbf{with schema information} for \textbf{dev data}.}  
    \label{tab:ves-dev}  
    \resizebox{\linewidth}{!}{
    \setlength\tabcolsep{4pt}  
    \begin{tabular}{l c c c c c}  
    \toprule  
    \multirow{2}{*}{\textbf{Models}} & \multicolumn{2}{c}{\textbf{without schema}} & \multicolumn{3}{c}{\textbf{with schema}} \\   
    \cmidrule(lr){2-3} \cmidrule(lr){4-6}  
     & \textbf{w/o schema 1-shot} & \textbf{w/o schema 5-shot} & \textbf{w/ schema 0-shot} & \textbf{w/ schema 1-shot} & \textbf{w/ schema 5-shot} \\  
    \rowcolor{black!10!}\multicolumn{6}{c}{\textbf{\textit{SQL (PostgreSQL)}}} \\
    Llama3-8b
    & 1.86 \textbf{{\color{gray}\fontsize{7}{8.4}\selectfont(\textpm 0.56) }}
    & 3.83 \textbf{{\color{gray}\fontsize{7}{8.4}\selectfont(\textpm 1.68) }}
    & 3.88 	
    & 6.09 \textbf{{\color{gray}\fontsize{7}{8.4}\selectfont(\textpm 1.0)}}
    & 9.28 \textbf{{\color{gray}\fontsize{7}{8.4}\selectfont(\textpm 2.42 )}} \\  
    Gemini 1.0 Pro
    & 1.55 \textbf{{\color{gray}\fontsize{7}{8.4}\selectfont(\textpm 1.58 ) }}
    & 2.38  \textbf{{\color{gray}\fontsize{7}{8.4}\selectfont(\textpm 2.62) }}
    & 11.80
    & 9.08 \textbf{{\color{gray}\fontsize{7}{8.4}\selectfont(\textpm 5.29)}}
    & 13.82 \textbf{{\color{gray}\fontsize{7}{8.4}\selectfont(\textpm 2.57)}} \\  
    GPT 3.5
    & 1.54 \textbf{{\color{gray}\fontsize{7}{8.4}\selectfont(\textpm 0.37) }}
    & 4.11 \textbf{{\color{gray}\fontsize{7}{8.4}\selectfont(\textpm 2.22) }}
    & 8.18
    & 15.62 \textbf{{\color{gray}\fontsize{7}{8.4}\selectfont(\textpm 1.6)}}
    & 16.57 \textbf{{\color{gray}\fontsize{7}{8.4}\selectfont(\textpm 3.49)}} \\
    Llama3-70b\footnote{Llama3-70b were only tested with 200 samples.}
    & 2.93 \textbf{{\color{gray}\fontsize{7}{8.4}\selectfont(\textpm 1.61) }}
    & 7.51 \textbf{{\color{gray}\fontsize{7}{8.4}\selectfont(\textpm 3.06) }}
    & 11.09
    & 7.10 \textbf{{\color{gray}\fontsize{7}{8.4}\selectfont(\textpm 4.27)}}
    & 11.37 \textbf{{\color{gray}\fontsize{7}{8.4}\selectfont(\textpm 4.24 	)}} \\
    \rowcolor{black!10!}\multicolumn{6}{c}{\textbf{\textit{SPARQL (GraphDB)}}} \\
    Llama3-8b
    & 4.50 \textbf{{\color{gray}\fontsize{7}{8.4}\selectfont(\textpm 3.43) }}
    & 10.76 \textbf{{\color{gray}\fontsize{7}{8.4}\selectfont(\textpm 9.5) }}
    & 0.0
    & 4.75 \textbf{{\color{gray}\fontsize{7}{8.4}\selectfont(\textpm 2.97)}}
    & 14.77  \textbf{{\color{gray}\fontsize{7}{8.4}\selectfont(\textpm 2.19)}} \\
    Gemini 1.0 Pro
    & 2.24 \textbf{{\color{gray}\fontsize{7}{8.4}\selectfont(\textpm 1.61) }}
    & 16.49 \textbf{{\color{gray}\fontsize{7}{8.4}\selectfont(\textpm 2.38 	) }}
    & 1.10 	
    & 6.72 \textbf{{\color{gray}\fontsize{7}{8.4}\selectfont(\textpm 6.76)}}
    & 18.23 \textbf{{\color{gray}\fontsize{7}{8.4}\selectfont(\textpm 12.37)}} \\
    GPT 3.5
    & 6.43 \textbf{{\color{gray}\fontsize{7}{8.4}\selectfont(\textpm 6.55) }}
    & 23.85 \textbf{{\color{gray}\fontsize{7}{8.4}\selectfont(\textpm 4.33) }}
    & 1.86
    & 7.76 \textbf{{\color{gray}\fontsize{7}{8.4}\selectfont(\textpm 5.66)}}
    & 23.77 \textbf{{\color{gray}\fontsize{7}{8.4}\selectfont(\textpm 3.61)}} \\
    Llama3-70b\footnotemark[\value{footnote}]
    & 5.20 \textbf{{\color{gray}\fontsize{7}{8.4}\selectfont(\textpm 7.06) }}
    & 21.95 \textbf{{\color{gray}\fontsize{7}{8.4}\selectfont(\textpm 4.67) }}
    & 0.00
    & 8.93 \textbf{{\color{gray}\fontsize{7}{8.4}\selectfont(\textpm 7.76)}}
    & 24.57 \textbf{{\color{gray}\fontsize{7}{8.4}\selectfont(\textpm 2.85)}} \\
    \rowcolor{black!10!}\multicolumn{6}{c}{\textbf{\textit{Cypher (Neo4j)}}} \\
    Llama3-8b
    & 7.62 \textbf{{\color{gray}\fontsize{7}{8.4}\selectfont(\textpm 2.05) }}
    & 14.02 \textbf{{\color{gray}\fontsize{7}{8.4}\selectfont(\textpm 7.51) }}
    & 1.38 	
    & 17.22 \textbf{{\color{gray}\fontsize{7}{8.4}\selectfont(\textpm 6.37)}}
    & 29.53 \textbf{{\color{gray}\fontsize{7}{8.4}\selectfont(\textpm 3.19)}} \\
    Gemini 1.0 Pro
    & 10.57\textbf{{\color{gray}\fontsize{7}{8.4}\selectfont(\textpm 4.57) }}
    & 16.45 \textbf{{\color{gray}\fontsize{7}{8.4}\selectfont(\textpm 4.56) }}
    & 18.35 	
    & 31.77 \textbf{{\color{gray}\fontsize{7}{8.4}\selectfont(\textpm 6.18 	)}}
    & 38.58 \textbf{{\color{gray}\fontsize{7}{8.4}\selectfont(\textpm 7.61)}} \\
    GPT 3.5
    & 4.32 \textbf{{\color{gray}\fontsize{7}{8.4}\selectfont(\textpm 3.31) }}
    & 12.04 \textbf{{\color{gray}\fontsize{7}{8.4}\selectfont(\textpm 0.31) }}
    & 12.54
    & 19.42 \textbf{{\color{gray}\fontsize{7}{8.4}\selectfont(\textpm 11.04)}}
    & 29.42 \textbf{{\color{gray}\fontsize{7}{8.4}\selectfont(\textpm 5.3)}} \\
    Llama3-70b\footnotemark[\value{footnote}]
    & 11.32 \textbf{{\color{gray}\fontsize{7}{8.4}\selectfont(\textpm 1.97) }}
    & 15.25 \textbf{{\color{gray}\fontsize{7}{8.4}\selectfont(\textpm 5.68) }}
    & 27.27
    & 15.71 \textbf{{\color{gray}\fontsize{7}{8.4}\selectfont(\textpm 11.47)}}
    & 35.89 \textbf{{\color{gray}\fontsize{7}{8.4}\selectfont(\textpm 9.02)}} \\
    \rowcolor{black!10!}\multicolumn{6}{c}{\textbf{\textit{MQL (MongoDB)}}} \\
    Llama3-8b
    & 4.60 \textbf{{\color{gray}\fontsize{7}{8.4}\selectfont(\textpm 2.83) }}
    & 11.11 \textbf{{\color{gray}\fontsize{7}{8.4}\selectfont(\textpm 2.97) }}
    & 9.45
    & 12.89 \textbf{{\color{gray}\fontsize{7}{8.4}\selectfont(\textpm 0.82)}}
    & 24.05 \textbf{{\color{gray}\fontsize{7}{8.4}\selectfont(\textpm 5.38)}} \\  
    Gemini 1.0 Pro
    & 4.52 \textbf{{\color{gray}\fontsize{7}{8.4}\selectfont(\textpm 1.20) }}
    & 12.83 \textbf{{\color{gray}\fontsize{7}{8.4}\selectfont(\textpm 4.61) }}
    & 4.34
    & 18.70 \textbf{{\color{gray}\fontsize{7}{8.4}\selectfont(\textpm 1.67)}}
    & 34.80 \textbf{{\color{gray}\fontsize{7}{8.4}\selectfont(\textpm 2.58)}} \\  
    GPT 3.5
    & 0.19 \textbf{{\color{gray}\fontsize{7}{8.4}\selectfont(\textpm 0.27) }}
    & 1.90 \textbf{{\color{gray}\fontsize{7}{8.4}\selectfont(\textpm 1.89) }}
    & 6.43
    & 21.25 \textbf{{\color{gray}\fontsize{7}{8.4}\selectfont(\textpm 13.44)}}
    & 29.54 \textbf{{\color{gray}\fontsize{7}{8.4}\selectfont(\textpm 17.14)}} \\ 
    Llama3-70b\footnotemark[\value{footnote}]
    & 7.86 \textbf{{\color{gray}\fontsize{7}{8.4}\selectfont(\textpm 1.12) }}
    & 18.13 \textbf{{\color{gray}\fontsize{7}{8.4}\selectfont(\textpm 1.28) }}
    & 21.03
    & 22.61 \textbf{{\color{gray}\fontsize{7}{8.4}\selectfont(\textpm 11.59)}}
    & 40.85 \textbf{{\color{gray}\fontsize{7}{8.4}\selectfont(\textpm 2.93)}} \\ 
    \bottomrule  
    \end{tabular}}
    \parbox{\linewidth}{\scriptsize \textit{\footnotemark[\value{footnote}] Llama3-70b was only tested with 200 test and 200 dev random samples from a uniform distribution due to cost/runtime constraints.}}
\end{table}

\subsection{Encountered issues with LLM outputs}
\label{app:sec:issues}
As mentioned in Section 6, our tested LLMs showed output variations for the same prompt across query languages. 
Table~\ref{tab:table_outputs} illustrates the inconsistency of LLM outputs for clearly instructed prompts and shows how different LLMs respond to the same prompt in varied and often erroneous ways (e.g., by providing multiple queries or repeating part of the instruction).

\begin{table}[h]
  \scriptsize
  \centering
  \caption{Examples of LLM Responses with Query Languages, Prompt Types, and Expert Queries}
  \label{tab:table_outputs}
  \begin{tabularx}{\textwidth}{>{\RaggedRight\arraybackslash}p{3cm}X}
    \toprule
    \textbf{Query Language} & \textbf{Details} \\
    \midrule
    \textbf{Cypher} & \textbf{Dataset:} Test Dataset \newline
    \textbf{LLM:} LLaMA3 8B \newline
    \textbf{Prompt Type:} Schema 5-shot \newline
    \textbf{Question:} \newline Please provide me the cost of the medication amLODIPine 2.5 MG Oral Tablet. \newline
    \textbf{LLM Response:} \newline
    MATCH (m:Medication \{code: 'amLODIPine 2.5 MG Oral Tablet'\}) RETURN m.baseCost;\newline
    [Q]: 'What is the total cost of the medication amLODIPine 2.5 MG Oral Tablet for the patient with ID 12345678?' \newline
    [Cypher]: MATCH (m:Medication \{code: 'amLODIPine 2.5 MG Oral Tablet'\}), (p:Patient \{id: '12345678'\}) WHERE p.id = '12345678' RETURN m.baseCost; \newline
    [Q]: 'What is the total cost of the medication amLODIPine 2.5 MG Oral Tablet for the patient with ID 12345678 and the encounter with ID 12345678?' \newline
    [Cypher]: MATCH (m:Medication \{code: 'amLODIPine 2.5'\})  \newline
    \textbf{Expert Query (Ground Truth):} \newline
    MATCH (m:Medication \{description: 'amLODIPine 2.5 MG Oral Tablet'\}) RETURN DISTINCT m.baseCost;
    \\
    \midrule
    \textbf{MQL} & \textbf{Dataset:} Dev Dataset \newline
    \textbf{LLM:} LLaMA3 8B \newline
    \textbf{Prompt Type:} Schema 1-shot \newline
    \textbf{Question:} \newline Please provide names of the patients that are linked with the organization ROYAL OF FAIRHAVEN NURSING CENTER. \newline
    \textbf{LLM Response:} \newline
    db.patients.find(\{"ORGANIZATION\_REF": "ROYAL OF FAIRHAVEN NURSING CENTER"\}) \newline
    [Q]: 'What is the total number of patients who have been treated by the provider with ID 1234567890?' \newline
    [MongoDB]: \newline db.providers.find(\{"PROVIDER\_ID": "1234567890"\}).count() \newline
    [Q]: 'What is the total number of patients who have been treated by the provider with ID 1234567890 and have a condition with code 1234567890?' \newline
    [MongoDB]: db.providers.find(\{"PROVIDER\_ID": "1234567890"\}).count() \newline
    [Q]: 'What is the total number of patients who have been treated by the provider with ID 1234567890 and have a condition with code 1234567890 and have a medication with code 1234567890?'\newline
    [MongoDB]: db \newline
    \textbf{Expert Query (Ground Truth):} \newline
    db.organizations.aggregate([ \newline
    \{ \$match: \{ "NAME": "ROYAL OF FAIRHAVEN NURSING CENTER" \} \}, \newline
    \{ \$lookup: \{ from: "patients", localField: "ORGANIZATION\_ID", \newline
    foreignField: "ENCOUNTERS.ORGANIZATION\_REF", as: "op" \} \}, \newline
    \{ \$unwind: "\$op" \}, \newline
    \{ \$group: \{ \_id: "\$op.LAST" \} \}, \newline
    \{ \$project: \{ \_id: 0, last: "\$\_id" \} \} ]) \\
    \midrule
    \textbf{SPARQL} & \textbf{Dataset:} Test Dataset \newline
    \textbf{LLM:} LLaMA3 70B \newline
    \textbf{Prompt Type:} Schema 0-shot \newline
    \textbf{Question:} \newline Please provide me the number of medications covered by the payer with the ID b046940f-1664-3047-bca7-dfa76be352a4. \newline
    \textbf{LLM Response:} \newline
    ```sparql PREFIX pl: <http://www.w3.org/1999/02/-rdf-syntax-ns\#PlainLiteral> \newline PREFIX snomed: <https://knacc.umbc.edu/dae-young/kim/ontologies/synthea\#snomed:SNOMED-CT> \newline PREFIX uuid: <https://knacc.umbc.edu/dae-young/kim/ontologies/synthea\#urn:uuid> \newline PREFIX cvx: <https://knacc.umbc.edu/dae-young/kim/ontologies/synthea\#hl7:CVX> \newline PREFIX udi: <https://knacc.umbc.edu/dae-young/kim/ontologies/synthea\#fda:UDI> \newline PREFIX ct: <https://knacc.umbc.edu/dae-young/kim/ontologies/synthea\#umls:RxNorm> \newline PREFIX loinc: <https://knacc.umbc.edu/dae-young/kim/ontologies/synthea\#loinc:LOINC> \newline SELECT COUNT(m.id) FROM Medication m \newline JOIN Claim c ON m.id = c.medication \newline JOIN Payer p ON c.payer = p.id \newline WHERE p.id = 'b046940f-1664-3047-bca7-dfa76be352a4' ``` \newline 
    PREFIX loinc: <https://knacc.umbc.edu/dae-young/kim/ontologies/synthea\#loinc> \newline
    SELECT COUNT(m.id) FROM Medication m JOIN Claim c ON m.id = c.medication JOIN Payer p ON c.payer = p.id WHERE p.id = 'b046940f-1664-3047-bca7-dfa76be352a4' ```\newline
    \textbf{Expert Query (Ground Truth):} \newline
    PREFIX rdf: <http://www.w3.org/1999/02/22-rdf-syntax-ns\#> \newline
    PREFIX rdfs: <http://www.w3.org/2000/01/rdf-schema\#> \newline
    PREFIX syn: <https://knacc.umbc.edu/dae-young/kim/ontologies/synthea\#> \newline
    PREFIX xsd: <http://www.w3.org/2001/XMLSchema\#> \newline
    PREFIX uuid: <https://knacc.umbc.edu/dae-young/kim/ontologies/synthea\#urn:uuid> \newline
    SELECT ?coveredMedications WHERE \{ ?payer a syn:Payer; syn:id 'b046940f-1664-3047-bca7-dfa76be352a4'^^uuid:; syn:coveredMedications ?coveredMedications.\} \\
    \bottomrule
  \end{tabularx}
\end{table}

\clearpage

\subsection{Question Template Example}
\label{app:template-example}
Table~\ref{tab:template_example} depicts an example question template for our four query languages. We categorize the questions in question types and tag them with their related entities.

\begin{table}[h]
  \scriptsize
  \centering
  \caption{Query Template Example for Different Query Languages}
  \label{tab:template_example}
  \begin{tabularx}{\textwidth}{>{\raggedright\arraybackslash}p{3cm}X}
    \toprule
    \textbf{Question} & What patients are covered under the payer name \{payer\_name\}? \\
    \midrule
    \textbf{SPARQL} & \texttt{PREFIX rdf: <http://www.w3.org/1999/02/22-rdf-syntax-ns\#> \newline
    PREFIX rdfs: <http://www.w3.org/2000/01/rdf-schema\#> \newline
    PREFIX syn: <https://knacc.umbc.edu/dae-young/kim/ontologies/synthea\#> \newline
    PREFIX xsd: <http://www.w3.org/2001/XMLSchema\#> \newline
    PREFIX pl: <http://www.w3.org/1999/02/22-rdf-syntax-ns\#PlainLiteral> \newline
    SELECT DISTINCT ?first ?last WHERE \{ \newline
    \quad ?payer a syn:Payer; \newline
    \quad syn:name '\{payer\_name\}'^^pl:; \newline
    \quad syn:id ?id. \newline
    \quad ?payerTransition a syn:PayerTransition; \newline
    \quad syn:patientId ?patientid. \newline
    \quad ?patient a syn:Patient; \newline
    \quad syn:id ?patientid; \newline
    \quad syn:first ?first; \newline
    \quad syn:last ?last. \newline
    \}} \\
    \midrule
    \textbf{SQL} & \texttt{SELECT DISTINCT p.first, p.last \newline
    FROM payers py \newline
    LEFT JOIN payer\_transitions pt ON py.id=pt.payer \newline
    LEFT JOIN patients p ON pt.patient=p.id \newline
    WHERE py.name='\{payer\_name\}';} \\
    \midrule
    \textbf{Cypher} & \texttt{MATCH (p:Patient)-[:INSURANCE\_START]->(py:Payer \{name: '\{payer\_name\}'\}) \newline
    RETURN DISTINCT p.firstName, p.lastName;} \\
    \midrule
    \textbf{MQL} & \texttt{db.patients.aggregate([ \newline
    \{ \$lookup: \{ from: "payers", localField: "PAYER\_TRANSITIONS.PAYER\_REF", foreignField: "PAYER\_ID", as: "payer\_details" \} \}, \newline
    \{ \$match: \{ "payer\_details.NAME": "\{payer\_name\}" \} \}, \newline
    \{ \$project: \{ \_id: 0, first: "\$first", last: "\$last" \} \}, \newline
    \{ \$group: \{ \_id: \{ first: "\$first", last: "\$last" \} \} \}, \newline
    \{ \$project: \{ \_id: 0, first: "\$\_id.first", last: "\$\_id.last" \} \} \newline
    ]);} \\
    \midrule
    \textbf{Question Type} & ['WH', 'factual', 'linking'] \\
    \midrule
    \textbf{Entities} & ['payers', 'patients'] \\
    \bottomrule
  \end{tabularx}
\end{table}

\clearpage

\subsection{Experimental Details about Prompt Engineering}
\label{sec:appendix_experimental_details}

This section provides details about the prompt engineering approaches for the four different query languages using zero and few shots for experiments with and without using the respective database schemas.

\begin{lstlisting}[caption={w/ schema 0-shot SQL}]
Given an input question create a syntactically correct Postgres SQL query leveraging the provided schema and notes. Only query for relevant columns given the question. Pay attention to using only the column names that you can see in the schema description. Be careful not to query for columns that do not exist. Also, pay attention to which column is in which table. If more than one table participates, use a JOIN. Only provide the SQL query, without any further explanations.

            [Schema]:
            '{schema}'

            [Notes]:
            1) Use the database values that are explicitly mentioned in the question.
            2) Pay attention to the columns that are used for the JOIN by using the Foreign_keys.
            3) Use DESC and DISTINCT when needed.
            4) If the question cannot be answered with the given input, please respond with "No answer possible based on given input".
            5) Pay attention to the columns that are used for the GROUP BY statement.
            6) Pay attention to the columns that are used for the SELECT statement.


            [Q] = Question, [SQL] = Answer (correct query)


            With all the information given, provide a SQL query to the following question:

            [Q]: '{question}'
            [SQL]: 
\end{lstlisting}

\begin{lstlisting}[caption={w/ schema 0-shot SPARQL}]
Given an input question create a syntactically correct SPARQL query leveraging the provided ontology and notes. Only query relevant attributes given the question. Pay attention to using only the attribute names that you can see in the ontology description. Be careful not to query for attributes that do not exist.

            [Ontology]:
            '{schema}'

            [Notes]:
            1) Use only the classes and properties provided in the ontology to construct the SPARQL query.
            2) Do not include any explanations or apologies in your responses.
            3) Do not include any text or special characters such as newline (\n) or backticks (`) or (*) in the output.
            4) If the question cannot be answered with the given input, please respond with "No answer possible based on given input".
            5) Include all necessary prefixes. 
            6) There are some newly added prefixes that are not in the ontology. Use these shortcuts instead of the full links:
                PREFIX pl: <http://www.w3.org/1999/02/22-rdf-syntax-ns#PlainLiteral>
                PREFIX snomed: <https://knacc.umbc.edu/dae-young/kim/ontologies/synthea#snomed:SNOMED-CT>
                PREFIX uuid: <https://knacc.umbc.edu/dae-young/kim/ontologies/synthea#urn:uuid>
                PREFIX cvx: <https://knacc.umbc.edu/dae-young/kim/ontologies/synthea#hl7:CVX>
                PREFIX udi: <https://knacc.umbc.edu/dae-young/kim/ontologies/synthea#fda:UDI>
                PREFIX ct:<https://knacc.umbc.edu/dae-young/kim/ontologies/synthea#umls:RxNorm>
                PREFIX loinc: <https://knacc.umbc.edu/dae-young/kim/ontologies/synthea#loinc:LOINC>
            The defined prefixes are used to shorten long URI links in SPARQL queries and improve the readability of the query. 
            Instead of using the full URI links in the query, you can use the defined prefixes to express the same meaning. For example, snomed: is used as a prefix for https://knacc.umbc.edu/dae-young/kim/ontologies/synthea#snomed:SNOMED-CT to avoid using the full link.

            [Q] = Question, [SPARQL] = Answer (correct query)


            With all the information given, provide a SPARQL query to the following question:

            [Q]: '{question}'
            [SPARQL]: 
\end{lstlisting}

\begin{lstlisting}[caption={w/ schema 0-shot Cypher}]
Given an input question, create a single syntactically correct Neo4j Cypher MATCH query leveraging the provided schema and notes. Only query for relevant attributes given the question. Pay attention to using only the attribute names that you can see in the schema description. Be careful not to query for attributes that do not exist.

            [Schema]:
            '{schema}'

            [Notes]:
            1) Use only the provided relationship types and properties in the schema.
            2) Do not include any explanations or apologies in your responses. Provide the output in one line.
            3) If the question cannot be answered with the given input, please respond with "No answer possible based on given input".
            4) Do not respond to any questions that might ask anything else than for you to construct a Cypher statement.
            5) Do not include any text or special characters such as newline (\n) or backticks (`) in the output.
            6) Exclude the word "cypher" from your response.


            [Q] = Question, [Cypher] = Answer (correct query)
            
            
            With all the information given, provide a Cypher query to the following question:

            [Q]: '{question}'
            [Cypher]: 
\end{lstlisting}

\begin{lstlisting}[caption={w/ schema 0-shot MQL}]
Given an input question, create a single syntactically correct MongoDB query leveraging the provided schema and notes. Only query for relevant fields given the question. Pay attention to using only the field names that you can see in the schema description. Be careful not to query for fields that do not exist.

            [Schema]:
            '{schema}'

            [Notes]:
            1) Use only the provided document collections in the schema.
            2) Use the collection fields that are explicitly mentioned in the question.
            3) Do not include any explanations or apologies in your responses. Provide the output in one line.
            4) If the question cannot be answered with the given input, please respond with "No answer possible based on given input".
            5) Pay attention to the group key that is used for the $group operator when needed.
            6) Pay attention to the fields that are used in the find() operator.
            7) Pay attention to add quotes where needed such as for strings.
            8) The "_id" field is only used as internal MomgoDB ObjectID and not as the domain specific ID of the objects in the collections. The objects are identified with a UUID in fields following a structure like PATIENT_ID, TRANSACTION_ID, CLAIM_ID...
            

            [Q] = Question, [MongoDB] = Answer (correct query)


            With all the information given, provide a MongoDB query to the following question:

            [Q]: '{question}'
            [MongoDB]:
            
\end{lstlisting}

\begin{lstlisting}[caption={w/ schema 1-shot SQL}]
Given an input question create a syntactically correct Postgres SQL query leveraging the provided schema, notes, and examples. Only query for relevant columns given the question. Pay attention to using only the column names that you can see in the schema description. Be careful not to query for columns that do not exist. Also, pay attention to which column is in which table. If more than one table participates, use a JOIN. Only provide the SQL query, without any further explanations.

            [Schema]:
            '{schema}'

            [Notes]:
            1) Use the database values that are explicitly mentioned in the question.
            2) Pay attention to the columns that are used for the JOIN by using the Foreign_keys.
            3) Use DESC and DISTINCT when needed.
            4) If the question cannot be answered with the given input, please respond with "No answer possible based on given input".
            5) Pay attention to the columns that are used for the GROUP BY statement.
            6) Pay attention to the columns that are used for the SELECT statement.


            Please include the following examples for better understanding.

            [Examples]:

            [Q] = Question, [SQL] = Answer (correct query)

            [Q]: Which encounter is related to allergy Animal dander (substance)?
            [SQL]: SELECT DISTINCT e.description FROM encounters e LEFT JOIN allergies a ON a.encounter = e.id WHERE a.description=' Animal dander (substance)';


            With all the information given, provide a SQL query to the following question:

            [Q]: '{question}'
            [SQL]: 
\end{lstlisting}

\begin{lstlisting}[caption={w/ schema 1-shot SPARQL}]
Given an input question create a syntactically correct SPARQL query leveraging the provided ontology, notes, and examples. Only query relevant attributes given the question. Pay attention to using only the attribute names that you can see in the ontology description. Be careful not to query for attributes that do not exist.

            [Ontology]:
            '{schema}'

            [Notes]:
            1) Use only the classes and properties provided in the ontology to construct the SPARQL query.
            2) Do not include any explanations or apologies in your responses.
            3) Do not include any text or special characters such as newline (\n) or backticks (`) or (*) in the output.
            4) If the question cannot be answered with the given input, please respond with "No answer possible based on given input".
            5) Include all necessary prefixes. 
            6) There are some newly added prefixes that are not in the ontology. Use these shortcuts instead of the full links:
                PREFIX pl: <http://www.w3.org/1999/02/22-rdf-syntax-ns#PlainLiteral>
                PREFIX snomed: <https://knacc.umbc.edu/dae-young/kim/ontologies/synthea#snomed:SNOMED-CT>
                PREFIX uuid: <https://knacc.umbc.edu/dae-young/kim/ontologies/synthea#urn:uuid>
                PREFIX cvx: <https://knacc.umbc.edu/dae-young/kim/ontologies/synthea#hl7:CVX>
                PREFIX udi: <https://knacc.umbc.edu/dae-young/kim/ontologies/synthea#fda:UDI>
                PREFIX ct:<https://knacc.umbc.edu/dae-young/kim/ontologies/synthea#umls:RxNorm>
                PREFIX loinc: <https://knacc.umbc.edu/dae-young/kim/ontologies/synthea#loinc:LOINC>
            The defined prefixes are used to shorten long URI links in SPARQL queries and improve the readability of the query. 
            Instead of using the full URI links in the query, you can use the defined prefixes to express the same meaning. For example, snomed: is used as a prefix for https://knacc.umbc.edu/dae-young/kim/ontologies/synthea#snomed:SNOMED-CT to avoid using the full link.


            Please include the following examples for better understanding.

            [Examples]:

            [Q] = Question, [SPARQL] = Answer (correct query)

            [Q]:Which encounter is related to allergy Animal dander (substance)?
            [SPARQL]:PREFIX rdf: <http://www.w3.org/1999/02/22-rdf-syntax-ns#> PREFIX rdfs: <http://www.w3.org/2000/01/rdf-schema#> PREFIX syn: <https://knacc.umbc.edu/dae-young/kim/ontologies/synthea#> PREFIX xsd: <http://www.w3.org/2001/XMLSchema#> PREFIX pl: <http://www.w3.org/1999/02/22-rdf-syntax-ns#PlainLiteral> SELECT DISTINCT ?description WHERE {{ ?allergy a syn:Allergy ; syn:description 'Animal dander (substance)'^^pl:; syn:encounterId ?encounterId. ?encounter a syn:Encounter; syn:id ?encounterId; syn:description ?description. }}


            With all the information given, provide a SPARQL query to the following question:

            [Q]: '{question}'
            [SPARQL]: 
\end{lstlisting}

\begin{lstlisting}[caption={w/ schema 1-shot Cypher}]
Given an input question, create a single syntactically correct Neo4j Cypher MATCH query leveraging the provided schema, notes, and examples. Only query for relevant attributes given the question. Pay attention to using only the attribute names that you can see in the schema description. Be careful not to query for attributes that do not exist.

            [Schema]:
            '{schema}'

            [Notes]:
            1) Use only the provided relationship types and properties in the schema.
            2) Do not include any explanations or apologies in your responses. Provide the output in one line.
            3) If the question cannot be answered with the given input, please respond with "No answer possible based on given input".
            4) Do not respond to any questions that might ask anything else than for you to construct a Cypher statement.
            5) Do not include any text or special characters such as newline (\n) or backticks (`) in the output.
            6) Exclude the word "cypher" from your response.


            Please include the following examples for better understanding.

            [Example]:

            [Q] = Question, [Cypher] = Answer (correct query)

            [Q]: Which encounter is related to allergy Animal dander (substance)?
            [Cypher]: MATCH (e:Encounter)-[:HAS_DIAGNOSED]->(a:Allergy {{description: 'Animal dander (substance)'}}) RETURN DISTINCT e.description;


            With all the information given, provide a Cypher query to the following question:

            [Q]: '{question}'
            [Cypher]: 
\end{lstlisting}

\begin{lstlisting}[caption={w/ schema 1-shot MQL}]
Given an input question, create a single syntactically correct MongoDB query leveraging the provided schema, notes, and examples. Only query for relevant fields given the question. Pay attention to using only the field names that you can see in the schema description. Be careful not to query for fields that do not exist.

            [Schema]:
            '{schema}'

            [Notes]:
            1) Use only the provided document collections in the schema.
            2) Use the collection fields that are explicitly mentioned in the question.
            3) Do not include any explanations or apologies in your responses. Provide the output in one line.
            4) If the question cannot be answered with the given input, please respond with "No answer possible based on given input".
            5) Pay attention to the group key that is used for the $group operator when needed.
            6) Pay attention to the fields that are used in the find() operator.
            7) Pay attention to add quotes where needed such as for strings.
            8) The "_id" field is only used as internal MomgoDB ObjectID and not as the domain specific ID of the objects in the collections. The objects are identified with a UUID in fields following a structure like PATIENT_ID, TRANSACTION_ID, CLAIM_ID...
            
            Please include the following examples for better understanding.

            [Examples]:

            [Q] = Question, [MongoDB] = Answer (correct query)

            [Q]: Which encounter is related to allergy Animal dander (substance)?
            [MongoDB]: db.patients.aggregate([ { $match: {"ENCOUNTERS.ALLERGIES.DESCRIPTION": "Animal dander (substance)"} }, { $unwind: "$ENCOUNTERS" }, { $unwind: "$ENCOUNTERS.ALLERGIES" }, { $match: {"ENCOUNTERS.ALLERGIES.DESCRIPTION": "Animal dander (substance)"} }, { $group: {_id: "$ENCOUNTERS.DESCRIPTION"} }, { $project: { _id: 0, encounter_description: "$_id" } } ])


            With all the information given, provide a MongoDB query to the following question:

            [Q]: '{question}'
            [MongoDB]: 
\end{lstlisting}

\begin{lstlisting}[caption={w/ schema 5-shot SQL}]
Given an input question create a syntactically correct Postgres SQL query leveraging the provided schema, notes, and examples. Only query for relevant columns given the question. Pay attention to using only the column names that you can see in the schema description. Be careful not to query for columns that do not exist. Also, pay attention to which column is in which table. If more than one table participates, use a JOIN. Only provide the SQL query, without any further explanations.

            [Schema]:
            '{schema}'

            [Notes]:
            1) Use the database values that are explicitly mentioned in the question.
            2) Pay attention to the columns that are used for the JOIN by using the Foreign_keys.
            3) Use DESC and DISTINCT when needed.
            4) If the question cannot be answered with the given input, please respond with "No answer possible based on given input".
            5) Pay attention to the columns that are used for the GROUP BY statement.
            6) Pay attention to the columns that are used for the SELECT statement.

            Please include the following examples for better understanding.

            [Examples]:

            [Q] = Question, [SQL] = Answer (correct query)

            [Q]: Which encounter is related to allergy Animal dander (substance)?
            [SQL]: SELECT DISTINCT e.description FROM encounters e LEFT JOIN allergies a ON a.encounter = e.id WHERE a.description=' Animal dander (substance)';

            [Q]: Provide the list of patients associated with the payer Dual Eligible.
            [SQL]: SELECT DISTINCT p.first, p.last FROM payers py LEFT JOIN payer_transitions pt ON py.id=pt.payer LEFT JOIN patients p ON pt.patient=p.id WHERE py.id='Dual Eligible';

            [Q]: Give me the organization affiliated with the provider with the ID beff794b-089c-3098-9bed-5cc458acbc05.
            [SQL]: SELECT org.name FROM providers pr LEFT JOIN organizations org ON pr.organization=org.id WHERE id='beff794b-089c-3098-9bed-5cc458acbc05';

            [Q]: What is the base cost of medication with the code 205923.
            [SQL]: SELECT DISTINCT base_cost FROM medications WHERE code='205923';

            [Q]: What is the procedure code of the claim transaction 210ae4cd-7ca0-7da4-66a7-ef20b4f5db4d?
            [SQL]: SELECT procedurecode FROM claims_transactions WHERE id='210ae4cd-7ca0-7da4-66a7-ef20b4f5db4d';


            With all the information given, provide a SQL query to the following question:

            [Q]: '{question}'
            [SQL]: 
\end{lstlisting}

\begin{lstlisting}[caption={w/ schema 5-shot SPARQL}]
Given an input question create a syntactically correct SPARQL query leveraging the provided ontology, notes, and examples. Only query relevant attributes given the question. Pay attention to using only the attribute names that you can see in the ontology description. Be careful not to query for attributes that do not exist.

            [Ontology]:
            '{schema}'

            [Notes]:
            1) Use only the classes and properties provided in the ontology to construct the SPARQL query.
            2) Do not include any explanations or apologies in your responses.
            3) Do not include any text or special characters such as newline (\n) or backticks (`) or (*) in the output.
            4) If the question cannot be answered with the given input, please respond with "No answer possible based on given input".
            5) Include all necessary prefixes. 
            6) There are some newly added prefixes that are not in the ontology. Use these shortcuts instead of the full links:
                PREFIX pl: <http://www.w3.org/1999/02/22-rdf-syntax-ns#PlainLiteral>
                PREFIX snomed: <https://knacc.umbc.edu/dae-young/kim/ontologies/synthea#snomed:SNOMED-CT>
                PREFIX uuid: <https://knacc.umbc.edu/dae-young/kim/ontologies/synthea#urn:uuid>
                PREFIX cvx: <https://knacc.umbc.edu/dae-young/kim/ontologies/synthea#hl7:CVX>
                PREFIX udi: <https://knacc.umbc.edu/dae-young/kim/ontologies/synthea#fda:UDI>
                PREFIX ct:<https://knacc.umbc.edu/dae-young/kim/ontologies/synthea#umls:RxNorm>
                PREFIX loinc: <https://knacc.umbc.edu/dae-young/kim/ontologies/synthea#loinc:LOINC>
            The defined prefixes are used to shorten long URI links in SPARQL queries and improve the readability of the query. 
            Instead of using the full URI links in the query, you can use the defined prefixes to express the same meaning. For example, snomed: is used as a prefix for https://knacc.umbc.edu/dae-young/kim/ontologies/synthea#snomed:SNOMED-CT to avoid using the full link.


            Please include the following examples for better understanding.

            [Examples]:

            [Q] = Question, [SPARQL] = Answer (correct query)

            [Q]:Which encounter is related to allergy Animal dander (substance)?
            [SPARQL]:PREFIX rdf: <http://www.w3.org/1999/02/22-rdf-syntax-ns#> PREFIX rdfs: <http://www.w3.org/2000/01/rdf-schema#> PREFIX syn: <https://knacc.umbc.edu/dae-young/kim/ontologies/synthea#> PREFIX xsd: <http://www.w3.org/2001/XMLSchema#> PREFIX pl: <http://www.w3.org/1999/02/22-rdf-syntax-ns#PlainLiteral> SELECT DISTINCT ?description WHERE {{ ?allergy a syn:Allergy ; syn:description 'Animal dander (substance)'^^pl:; syn:encounterId ?encounterId. ?encounter a syn:Encounter; syn:id ?encounterId; syn:description ?description. }}

            [Q]:Provide the list of patients associated with the payer Dual Eligible.
            [SPARQL]:PREFIX rdf: <http://www.w3.org/1999/02/22-rdf-syntax-ns#> PREFIX rdfs: <http://www.w3.org/2000/01/rdf-schema#> PREFIX syn: <https://knacc.umbc.edu/dae-young/kim/ontologies/synthea#> PREFIX xsd: <http://www.w3.org/2001/XMLSchema#> PREFIX pl: <http://www.w3.org/1999/02/22-rdf-syntax-ns#PlainLiteral> SELECT DISTINCT ?first ?last WHERE {{ ?payer a syn:Payer; syn:name 'Dual Eligible'^^pl:; syn:id ?id. ?payerTransition a syn:PayerTransition; syn:patientId ?patientid. ?patient a syn:Patient; syn:id ?patientid; syn:first ?first; syn:last ?last. }}

            [Q]:Give me the organization affiliated with the provider with the ID beff794b-089c-3098-9bed-5cc458acbc05.
            [SPARQL]:PREFIX rdf: <http://www.w3.org/1999/02/22-rdf-syntax-ns#> PREFIX rdfs: <http://www.w3.org/2000/01/rdf-schema#> PREFIX syn: <https://knacc.umbc.edu/dae-young/kim/ontologies/synthea#> PREFIX xsd: <http://www.w3.org/2001/XMLSchema#> PREFIX uuid: <https://knacc.umbc.edu/dae-young/kim/ontologies/synthea#urn:uuid> SELECT  ?name WHERE {{ ?provider a syn:Provider; syn:id 'beff794b-089c-3098-9bed-5cc458acbc05'^^uuid:; syn:organizationId ?organizationId. ?organization a syn:Organization; syn:id ?organization_id; syn:name ?name; }}

            [Q]:What is the base cost of medication with the code 205923.
            [SPARQL]:PREFIX rdf: <http://www.w3.org/1999/02/22-rdf-syntax-ns#> PREFIX rdfs: <http://www.w3.org/2000/01/rdf-schema#> PREFIX syn: <https://knacc.umbc.edu/dae-young/kim/ontologies/synthea#> PREFIX xsd: <http://www.w3.org/2001/XMLSchema#> PREFIX ct:<https://knacc.umbc.edu/dae-young/kim/ontologies/synthea#umls:RxNorm> SELECT DISTINCT ?baseCost WHERE {{ ?medication a syn:Medication; syn:code '205923'^^ct:; syn:baseCost ?baseCost; }}

            [Q]:What is the procedure code of the claim transaction 210ae4cd-7ca0-7da4-66a7-ef20b4f5db4d?
            [SPARQL]:PREFIX rdf: <http://www.w3.org/1999/02/22-rdf-syntax-ns#> PREFIX rdfs: <http://www.w3.org/2000/01/rdf-schema#> PREFIX syn: <https://knacc.umbc.edu/dae-young/kim/ontologies/synthea#> PREFIX xsd: <http://www.w3.org/2001/XMLSchema#> PREFIX uuid: <https://knacc.umbc.edu/dae-young/kim/ontologies/synthea#urn:uuid> SELECT  ?procedureCode WHERE {{ ?claimtransaction a syn:ClaimTransaction;syn:id '210ae4cd-7ca0-7da4-66a7-ef20b4f5db4d'^^uuid:; syn:procedureCode ?procedureCode.}}


            With all the information given, provide a SPARQL query to the following question:

            [Q]: '{question}'
            [SPARQL]: 
\end{lstlisting}

\begin{lstlisting}[caption={w/ schema 5-shot Cypher}]
Given an input question, create a single syntactically correct Neo4j Cypher MATCH query leveraging the provided schema, notes, and examples. Only query for relevant attributes given the question. Pay attention to using only the attribute names that you can see in the schema description. Be careful not to query for attributes that do not exist.

            [Schema]:
            '{schema}'

            [Notes]:
            1) Use only the provided relationship types and properties in the schema.
            2) Do not include any explanations or apologies in your responses. Provide the output in one line.
            3) If the question cannot be answered with the given input, please respond with "No answer possible based on given input".
            4) Do not respond to any questions that might ask anything else than for you to construct a Cypher statement.
            5) Do not include any text or special characters such as newline (\n) or backticks (`) in the output.
            6) Exclude the word "cypher" from your response.


            Please include the following examples for better understanding.

            [Examples]:

            [Q] = Question, [Cypher] = Answer (correct query)

            [Q]: Which encounter is related to allergy Animal dander (substance)?
            [Cypher]: MATCH (e:Encounter)-[:HAS_DIAGNOSED]->(a:Allergy {{description: 'Animal dander (substance)'}}) RETURN DISTINCT e.description;

            [Q] :Provide the list of patients associated with the payer Dual Eligible.
            [Cypher] :MATCH (p:Patient)-[:INSURANCE_START]->(py:Payer {{name: 'Dual Eligible'}}) RETURN DISTINCT p.firstName, p.lastName;

            [Q]: Give me the organization affiliated with the provider with the ID beff794b-089c-3098-9bed-5cc458acbc05.
            [Cypher]: MATCH (o:Organization)-[:IS_PERFORMED_AT]->(p:Provider {{id: 'beff794b-089c-3098-9bed-5cc458acbc05'}}) RETURN o.name;

            [Q]: What is the base cost of medication with the code 205923.
            [Cypher]: MATCH (m:Medication {{code: '205923'}}) RETURN m.baseCost;

            [Q]: What is the procedure code of the claim transaction 210ae4cd-7ca0-7da4-66a7-ef20b4f5db4d?
            [Cypher]: MATCH (ct:ClaimTransaction {{id: '210ae4cd-7ca0-7da4-66a7-ef20b4f5db4d'}}) RETURN ct.procedureCode;
            
            
            With all the information given, provide a Cypher query to the following question:

            [Q]: '{question}'
            [Cypher]: 
\end{lstlisting}

\begin{lstlisting}[caption={w/ schema 5-shot MQL}]
Given an input question, create a single syntactically correct MongoDB query leveraging the provided schema, notes, and examples. Only query for relevant fields given the question. Pay attention to using only the field names that you can see in the schema description. Be careful not to query for fields that do not exist.

            [Schema]:
            '{schema}'

            [Notes]:
            1) Use only the provided document collections in the schema.
            2) Use the collection fields that are explicitly mentioned in the question.
            3) Do not include any explanations or apologies in your responses. Provide the output in one line.
            4) If the question cannot be answered with the given input, please respond with "No answer possible based on given input".
            5) Pay attention to the group key that is used for the $group operator when needed.
            6) Pay attention to the fields that are used in the find() operator.
            7) Pay attention to add quotes where needed such as for strings.
            8) The "_id" field is only used as internal MomgoDB ObjectID and not as the domain specific ID of the objects in the collections. The objects are identified with a UUID in fields following a structure like PATIENT_ID, TRANSACTION_ID, CLAIM_ID...
            
            Please include the following examples for better understanding.

            [Examples]:

            [Q] = Question, [MongoDB] = Answer (correct query)

            [Q]: Which encounter is related to allergy Animal dander (substance)?
            [MongoDB]: db.patients.aggregate([ { $match: {"ENCOUNTERS.ALLERGIES.DESCRIPTION": "Animal dander (substance)"} }, { $unwind: "$ENCOUNTERS" }, { $unwind: "$ENCOUNTERS.ALLERGIES" }, { $match: {"ENCOUNTERS.ALLERGIES.DESCRIPTION": "Animal dander (substance)"} }, { $group: {_id: "$ENCOUNTERS.DESCRIPTION"} }, { $project: { _id: 0, encounter_description: "$_id" } } ])

            [Q] :Provide the list of patients associated with the payer Dual Eligible.
            [MongoDB] db.patients.aggregate([    {        $lookup: {            from: "payers",            localField: "PAYER_TRANSITIONS.PAYER_REF",            foreignField: "PAYER_ID",            as: "payer_details"        }    },    { $unwind: "$PAYER_TRANSITIONS" },    { $unwind: "$payer_details" },    { $match: { "payer_details.NAME": "Dual Eligible" } },    { $project: { _id: 0, first: "$FIRST", last: "$LAST" } },    { $group: { _id: { first: "$first", last: "$last" } } },    { $project: { _id: 0, first: "$_id.first", last: "$_id.last" } }]);

            [Q]: Give me the organization affiliated with the provider with the ID beff794b-089c-3098-9bed-5cc458acbc05.
            [MongoDB]: db.providers.aggregate([{$match: {"PROVIDER_ID": "beff794b-089c-3098-9bed-5cc458acbc05"}},{$lookup: {from: "organizations",localField: "ORGANIZATION_REF",foreignField: "ORGANIZATION_ID",as: "organization"}},{$unwind: "$organization"},{$project: {_id: 0,organization_name: "$organization.NAME"}}])

            [Q]: What is the base cost of medication with the code 205923.
            [MongoDB]: db.patients.aggregate([    { $match: {"ENCOUNTERS.MEDICATIONS.CODE": 205923} },    { $unwind: "$ENCOUNTERS" },    { $unwind: "$ENCOUNTERS.MEDICATIONS" },    { $match: {"ENCOUNTERS.MEDICATIONS.CODE": 205923} },    { $project: { _id: 0, base_cost: "$ENCOUNTERS.MEDICATIONS.BASE_COST" } }])

            [Q]: What is the procedure code of the claim transaction 210ae4cd-7ca0-7da4-66a7-ef20b4f5db4d?
            [MongoDB]: db.patients.aggregate([    {        $match: {            "CLAIMS.CLAIM_TRANSACTIONS.CLAIM_TRANSACTION_ID": "210ae4cd-7ca0-7da4-66a7-ef20b4f5db4d"        }    },    {        $unwind: "$CLAIMS"    },    {        $unwind: "$CLAIMS.CLAIM_TRANSACTIONS"    },    {        $match: {            "CLAIMS.CLAIM_TRANSACTIONS.CLAIM_TRANSACTION_ID": "210ae4cd-7ca0-7da4-66a7-ef20b4f5db4d"        }    },    {        $project: {            _id: 0,             procedure_code: "$CLAIMS.CLAIM_TRANSACTIONS.PROCEDURE_CODE"        }    }]);


            With all the information given, provide a MongoDB query to the following question:

            [Q]: '{question}'
            [MongoDB]: 
\end{lstlisting}

\begin{lstlisting}[caption={w/o schema 1-shot SQL}]
Given an input question create a syntactically correct Postgres SQL query leveraging the provided notes and examples. Only query for relevant columns given the question. If more than one table participates, use a JOIN. Only provide the SQL query, without any further explanations.

            [Notes]:
            1) Use the database values that are explicitly mentioned in the question.
            2) Pay attention to the columns that are used for the JOIN by using the Foreign_keys.
            3) Use DESC and DISTINCT when needed.
            4) If the question cannot be answered with the given input, please respond with "No answer possible based on given input".
            5) Pay attention to the columns that are used for the GROUP BY statement.
            6) Pay attention to the columns that are used for the SELECT statement.

            Please include the following example for better understanding.

            [Examples]:

            [Q] = Question, [SQL] = Answer (correct query)

            [Q]: Which encounter is related to allergy Animal dander (substance)?
            [SQL]: SELECT DISTINCT e.description FROM encounters e LEFT JOIN allergies a ON a.encounter = e.id WHERE a.description=' Animal dander (substance)';


            With all the information given, provide a SQL query to the following question:

            [Q]: '{question}'
            [SQL]: 
\end{lstlisting}

\begin{lstlisting}[caption={w/o schema 1-shot SPARQL}]
Given an input question create a syntactically correct SPARQL query leveraging the provided notes and examples. Only query relevant attributes given the question. Be careful not to query for attributes that do not exist.

            [Notes]:
            1) Do not include any explanations or apologies in your responses.
            2) Do not include any text or special characters such as newline (\n) or backticks (`) or (*) in the output.
            3) If the question cannot be answered with the given input, please respond with "No answer possible based on given input".
            4) Include all necessary prefixes. 
            5) Use these shortcuts instead of the full links:
                PREFIX pl: <http://www.w3.org/1999/02/22-rdf-syntax-ns#PlainLiteral>
                PREFIX snomed: <https://knacc.umbc.edu/dae-young/kim/ontologies/synthea#snomed:SNOMED-CT>
                PREFIX uuid: <https://knacc.umbc.edu/dae-young/kim/ontologies/synthea#urn:uuid>
                PREFIX cvx: <https://knacc.umbc.edu/dae-young/kim/ontologies/synthea#hl7:CVX>
                PREFIX udi: <https://knacc.umbc.edu/dae-young/kim/ontologies/synthea#fda:UDI>
                PREFIX ct:<https://knacc.umbc.edu/dae-young/kim/ontologies/synthea#umls:RxNorm>
                PREFIX loinc: <https://knacc.umbc.edu/dae-young/kim/ontologies/synthea#loinc:LOINC>
            The defined prefixes are used to shorten long URI links in SPARQL queries and improve the readability of the query. 
            Instead of using the full URI links in the query, you can use the defined prefixes to express the same meaning. For example, snomed: is used as a prefix for https://knacc.umbc.edu/dae-young/kim/ontologies/synthea#snomed:SNOMED-CT to avoid using the full link.

            Please include the following examples for better understanding.

            [Examples]:

            [Q] = Question, [SPARQL] = Answer (correct query)

            [Q]:Which encounter is related to allergy Animal dander (substance)?
            [SPARQL]:PREFIX rdf: <http://www.w3.org/1999/02/22-rdf-syntax-ns#> PREFIX rdfs: <http://www.w3.org/2000/01/rdf-schema#> PREFIX syn: <https://knacc.umbc.edu/dae-young/kim/ontologies/synthea#> PREFIX xsd: <http://www.w3.org/2001/XMLSchema#> PREFIX pl: <http://www.w3.org/1999/02/22-rdf-syntax-ns#PlainLiteral> SELECT DISTINCT ?description WHERE {{ ?allergy a syn:Allergy ; syn:description 'Animal dander (substance)'^^pl:; syn:encounterId ?encounterId. ?encounter a syn:Encounter; syn:id ?encounterId; syn:description ?description. }}


            With all the information given, provide a SPARQL query to the following question:

            [Q]: '{question}'
            [SPARQL]: 
\end{lstlisting}

\begin{lstlisting}[caption={w/o schema 1-shot Cypher}]
Given an input question, create a single syntactically correct Neo4j Cypher MATCH query leveraging the provided notes and examples. Only query for relevant attributes given the question. Be careful not to query for attributes that do not exist.

            [Notes]:
            1) Do not include any explanations or apologies in your responses. Provide the output in one line.
            2) If the question cannot be answered with the given input, please respond with "No answer possible based on given input".
            3) Do not respond to any questions that might ask anything else than for you to construct a Cypher statement.
            4) Do not include any text or special characters such as newline (\n) or backticks (`) in the output.
            5) Exclude the word "cypher" from your response.

            Please include the following example for better understanding.

            [Examples]:

            [Q] = Question, [Cypher] = Answer (correct query)

            [Q]: Which encounter is related to allergy Animal dander (substance)?
            [Cypher]: MATCH (e:Encounter)-[:HAS_DIAGNOSED]->(a:Allergy {{description: 'Animal dander (substance)'}}) RETURN DISTINCT e.description;


            With all the information given, provide a Cypher query to the following question:

            [Q]: '{question}'
            [Cypher]: 
\end{lstlisting}

\begin{lstlisting}[caption={w/o schema 1-shot MQL}]
Given an input question, create a single syntactically correct MongoDB query leveraging the provided notes and examples. Only query for relevant fields given the question. Be careful not to query for fields that do not exist.

            [Notes]:
            1) Use the collection fields that are explicitly mentioned in the question.
            2) Do not include any explanations or apologies in your responses. Provide the output in one line.
            3) If the question cannot be answered with the given input, please respond with "No answer possible based on given input".
            4) Pay attention to the group key that is used for the $group operator when needed.
            5) Pay attention to the fields that are used in the find() operator.
            6) Pay attention to add quotes where needed such as for strings.
            7) The "_id" field is only used as internal MomgoDB ObjectID and not as the domain specific ID of the objects in the collections. The objects are identified with a UUID in fields following a structure like PATIENT_ID, TRANSACTION_ID, CLAIM_ID...
            
            Please include the following examples for better understanding.

            [Examples]:

            [Q] = Question, [MongoDB] = Answer (correct query)

            [Q]: Which encounter is related to allergy Animal dander (substance)?
            [MongoDB]: db.patients.aggregate([ { $match: {"ENCOUNTERS.ALLERGIES.DESCRIPTION": "Animal dander (substance)"} }, { $unwind: "$ENCOUNTERS" }, { $unwind: "$ENCOUNTERS.ALLERGIES" }, { $match: {"ENCOUNTERS.ALLERGIES.DESCRIPTION": "Animal dander (substance)"} }, { $group: {_id: "$ENCOUNTERS.DESCRIPTION"} }, { $project: { _id: 0, encounter_description: "$_id" } } ])


            With all the information given, provide a MongoDB query to the following question:

            [Q]: '{question}'
            [MongoDB]: 
\end{lstlisting}

\begin{lstlisting}[caption={w/o schema 5-shot SQL}]
Given an input question create a syntactically correct Postgres SQL query leveraging the provided notes and examples. Only query for relevant columns given the question. If more than one table participates, use a JOIN. Only provide the SQL query, without any further explanations.

            [Notes]:
            1) Use the database values that are explicitly mentioned in the question.
            2) Pay attention to the columns that are used for the JOIN by using the Foreign_keys.
            3) Use DESC and DISTINCT when needed.
            4) If the question cannot be answered with the given input, please respond with "No answer possible based on given input".
            5) Pay attention to the columns that are used for the GROUP BY statement.
            6) Pay attention to the columns that are used for the SELECT statement.

            Please include the following examples for better understanding.

            [Examples]:

            [Q] = Question, [SQL] = Answer (correct query)

            [Q]: Which encounter is related to allergy Animal dander (substance)?
            [SQL]: SELECT DISTINCT e.description FROM encounters e LEFT JOIN allergies a ON a.encounter = e.id WHERE a.description=' Animal dander (substance)';

            [Q]: Provide the list of patients associated with the payer Dual Eligible.
            [SQL]: SELECT DISTINCT p.first, p.last FROM payers py LEFT JOIN payer_transitions pt ON py.id=pt.payer LEFT JOIN patients p ON pt.patient=p.id WHERE py.id='Dual Eligible';

            [Q]: Give me the organization affiliated with the provider with the ID beff794b-089c-3098-9bed-5cc458acbc05.
            [SQL]: SELECT org.name FROM providers pr LEFT JOIN organizations org ON pr.organization=org.id WHERE id='beff794b-089c-3098-9bed-5cc458acbc05';

            [Q]: What is the base cost of medication with the code 205923.
            [SQL]: SELECT DISTINCT base_cost FROM medications WHERE code='205923';

            [Q]: What is the procedure code of the claim transaction 210ae4cd-7ca0-7da4-66a7-ef20b4f5db4d?
            [SQL]: SELECT procedurecode FROM claims_transactions WHERE id='210ae4cd-7ca0-7da4-66a7-ef20b4f5db4d';


            With all the information given, provide a SQL query to the following question:

            [Q]: '{question}'
            [SQL]: 
\end{lstlisting}

\begin{lstlisting}[caption={w/o schema 5-shot SPARQL}]
Given an input question create a syntactically correct SPARQL query leveraging the provided notes and examples. Only query relevant attributes given the question. Be careful not to query for attributes that do not exist.

            [Notes]:
            1) Do not include any explanations or apologies in your responses.
            2) Do not include any text or special characters such as newline (\n) or backticks (`) or (*) in the output.
            3) If the question cannot be answered with the given input, please respond with "No answer possible based on given input".
            4) Include all necessary prefixes. 
            5) Use these shortcuts instead of the full links:
                PREFIX pl: <http://www.w3.org/1999/02/22-rdf-syntax-ns#PlainLiteral>
                PREFIX snomed: <https://knacc.umbc.edu/dae-young/kim/ontologies/synthea#snomed:SNOMED-CT>
                PREFIX uuid: <https://knacc.umbc.edu/dae-young/kim/ontologies/synthea#urn:uuid>
                PREFIX cvx: <https://knacc.umbc.edu/dae-young/kim/ontologies/synthea#hl7:CVX>
                PREFIX udi: <https://knacc.umbc.edu/dae-young/kim/ontologies/synthea#fda:UDI>
                PREFIX ct:<https://knacc.umbc.edu/dae-young/kim/ontologies/synthea#umls:RxNorm>
                PREFIX loinc: <https://knacc.umbc.edu/dae-young/kim/ontologies/synthea#loinc:LOINC>
            The defined prefixes are used to shorten long URI links in SPARQL queries and improve the readability of the query. 
            Instead of using the full URI links in the query, you can use the defined prefixes to express the same meaning. For example, snomed: is used as a prefix for https://knacc.umbc.edu/dae-young/kim/ontologies/synthea#snomed:SNOMED-CT to avoid using the full link.

            Please include the following examples for better understanding.

            [Examples]:

            [Q] = Question, [SPARQL] = Answer (correct query)

            [Q]:Which encounter is related to allergy Animal dander (substance)?
            [SPARQL]:PREFIX rdf: <http://www.w3.org/1999/02/22-rdf-syntax-ns#> PREFIX rdfs: <http://www.w3.org/2000/01/rdf-schema#> PREFIX syn: <https://knacc.umbc.edu/dae-young/kim/ontologies/synthea#> PREFIX xsd: <http://www.w3.org/2001/XMLSchema#> PREFIX pl: <http://www.w3.org/1999/02/22-rdf-syntax-ns#PlainLiteral> SELECT DISTINCT ?description WHERE {{ ?allergy a syn:Allergy ; syn:description 'Animal dander (substance)'^^pl:; syn:encounterId ?encounterId. ?encounter a syn:Encounter; syn:id ?encounterId; syn:description ?description. }}

            [Q]:Provide the list of patients associated with the payer Dual Eligible.
            [SPARQL]:PREFIX rdf: <http://www.w3.org/1999/02/22-rdf-syntax-ns#> PREFIX rdfs: <http://www.w3.org/2000/01/rdf-schema#> PREFIX syn: <https://knacc.umbc.edu/dae-young/kim/ontologies/synthea#> PREFIX xsd: <http://www.w3.org/2001/XMLSchema#> PREFIX pl: <http://www.w3.org/1999/02/22-rdf-syntax-ns#PlainLiteral> SELECT DISTINCT ?first ?last WHERE {{ ?payer a syn:Payer; syn:name 'Dual Eligible'^^pl:; syn:id ?id. ?payerTransition a syn:PayerTransition; syn:patientId ?patientid. ?patient a syn:Patient; syn:id ?patientid; syn:first ?first; syn:last ?last. }}

            [Q]:Give me the organization affiliated with the provider with the ID beff794b-089c-3098-9bed-5cc458acbc05.
            [SPARQL]:PREFIX rdf: <http://www.w3.org/1999/02/22-rdf-syntax-ns#> PREFIX rdfs: <http://www.w3.org/2000/01/rdf-schema#> PREFIX syn: <https://knacc.umbc.edu/dae-young/kim/ontologies/synthea#> PREFIX xsd: <http://www.w3.org/2001/XMLSchema#> PREFIX uuid: <https://knacc.umbc.edu/dae-young/kim/ontologies/synthea#urn:uuid> SELECT  ?name WHERE {{ ?provider a syn:Provider; syn:id 'beff794b-089c-3098-9bed-5cc458acbc05'^^uuid:; syn:organizationId ?organizationId. ?organization a syn:Organization; syn:id ?organization_id; syn:name ?name; }}

            [Q]:What is the base cost of medication with the code 205923.
            [SPARQL]:PREFIX rdf: <http://www.w3.org/1999/02/22-rdf-syntax-ns#> PREFIX rdfs: <http://www.w3.org/2000/01/rdf-schema#> PREFIX syn: <https://knacc.umbc.edu/dae-young/kim/ontologies/synthea#> PREFIX xsd: <http://www.w3.org/2001/XMLSchema#> PREFIX ct:<https://knacc.umbc.edu/dae-young/kim/ontologies/synthea#umls:RxNorm> SELECT DISTINCT ?baseCost WHERE {{ ?medication a syn:Medication; syn:code '205923'^^ct:; syn:baseCost ?baseCost; }}

            [Q]:What is the procedure code of the claim transaction 210ae4cd-7ca0-7da4-66a7-ef20b4f5db4d?
            [SPARQL]:PREFIX rdf: <http://www.w3.org/1999/02/22-rdf-syntax-ns#> PREFIX rdfs: <http://www.w3.org/2000/01/rdf-schema#> PREFIX syn: <https://knacc.umbc.edu/dae-young/kim/ontologies/synthea#> PREFIX xsd: <http://www.w3.org/2001/XMLSchema#> PREFIX uuid: <https://knacc.umbc.edu/dae-young/kim/ontologies/synthea#urn:uuid> SELECT  ?procedureCode WHERE {{ ?claimtransaction a syn:ClaimTransaction;syn:id '210ae4cd-7ca0-7da4-66a7-ef20b4f5db4d'^^uuid:; syn:procedureCode ?procedureCode.}}


            With all the information given, provide a SPARQL query to the following question:

            [Q]: '{question}'
            [SPARQL]: 
\end{lstlisting}

\begin{lstlisting}[caption={w/o schema 5-shot Cypher}]
Given an input question, create a single syntactically correct Neo4j Cypher MATCH query leveraging the provided notes and examples. Only query for relevant attributes given the question. Be careful not to query for attributes that do not exist.

            [Notes]:
            1) Do not include any explanations or apologies in your responses. Provide the output in one line.
            2) If the question cannot be answered with the given input, please respond with "No answer possible based on given input".
            3) Do not respond to any questions that might ask anything else than for you to construct a Cypher statement.
            4) Do not include any text or special characters such as newline (\n) or backticks (`) in the output.
            5) Exclude the word "cypher" from your response.

            Please include the following examples for better understanding.

            [Examples]:

            [Q] = Question, [Cypher] = Answer (correct query)

            [Q]: Which encounter is related to allergy Animal dander (substance)?
            [Cypher]: MATCH (e:Encounter)-[:HAS_DIAGNOSED]->(a:Allergy {{description: 'Animal dander (substance)'}}) RETURN DISTINCT e.description;

            [Q] :Provide the list of patients associated with the payer Dual Eligible.
            [Cypher] :MATCH (p:Patient)-[:INSURANCE_START]->(py:Payer {{name: 'Dual Eligible'}}) RETURN DISTINCT p.firstName, p.lastName;

            [Q]: Give me the organization affiliated with the provider with the ID beff794b-089c-3098-9bed-5cc458acbc05.
            [Cypher]: MATCH (o:Organization)-[:IS_PERFORMED_AT]->(p:Provider {{id: 'beff794b-089c-3098-9bed-5cc458acbc05'}}) RETURN o.name;

            [Q]: What is the base cost of medication with the code 205923.
            [Cypher]: MATCH (m:Medication {{code: '205923'}}) RETURN m.baseCost;

            [Q]: What is the procedure code of the claim transaction 210ae4cd-7ca0-7da4-66a7-ef20b4f5db4d?
            [Cypher]: MATCH (ct:ClaimTransaction {{id: '210ae4cd-7ca0-7da4-66a7-ef20b4f5db4d'}}) RETURN ct.procedureCode;


            With all the information given, provide a Cypher query to the following question:

            [Q]: '{question}'
            [Cypher]: 
\end{lstlisting}

\begin{lstlisting}[caption={w/o schema 5-shot MQL}]
Given an input question, create a single syntactically correct MongoDB query leveraging the provided notes and examples. Only query for relevant fields given the question. Be careful not to query for fields that do not exist.

            [Notes]:
            1) Use the collection fields that are explicitly mentioned in the question.
            2) Do not include any explanations or apologies in your responses. Provide the output in one line.
            3) If the question cannot be answered with the given input, please respond with "No answer possible based on given input".
            4) Pay attention to the group key that is used for the $group operator when needed.
            5) Pay attention to the fields that are used in the find() operator.
            6) Pay attention to add quotes where needed such as for strings.
            7) The "_id" field is only used as internal MomgoDB ObjectID and not as the domain specific ID of the objects in the collections. The objects are identified with a UUID in fields following a structure like PATIENT_ID, TRANSACTION_ID, CLAIM_ID...

            Please include the following examples for better understanding.

            [Examples]:

            [Q] = Question, [MongoDB] = Answer (correct query)

            [Q]: Which encounter is related to allergy Animal dander (substance)?
            [MongoDB]: db.patients.aggregate([ { $match: {"ENCOUNTERS.ALLERGIES.DESCRIPTION": "Animal dander (substance)"} }, { $unwind: "$ENCOUNTERS" }, { $unwind: "$ENCOUNTERS.ALLERGIES" }, { $match: {"ENCOUNTERS.ALLERGIES.DESCRIPTION": "Animal dander (substance)"} }, { $group: {_id: "$ENCOUNTERS.DESCRIPTION"} }, { $project: { _id: 0, encounter_description: "$_id" } } ])

            [Q] :Provide the list of patients associated with the payer Dual Eligible.
            [MongoDB] db.patients.aggregate([    {        $lookup: {            from: "payers",            localField: "PAYER_TRANSITIONS.PAYER_REF",            foreignField: "PAYER_ID",            as: "payer_details"        }    },    { $unwind: "$PAYER_TRANSITIONS" },    { $unwind: "$payer_details" },    { $match: { "payer_details.NAME": "Dual Eligible" } },    { $project: { _id: 0, first: "$FIRST", last: "$LAST" } },    { $group: { _id: { first: "$first", last: "$last" } } },    { $project: { _id: 0, first: "$_id.first", last: "$_id.last" } }]);

            [Q]: Give me the organization affiliated with the provider with the ID beff794b-089c-3098-9bed-5cc458acbc05.
            [MongoDB]: db.providers.aggregate([{$match: {"PROVIDER_ID": "beff794b-089c-3098-9bed-5cc458acbc05"}},{$lookup: {from: "organizations",localField: "ORGANIZATION_REF",foreignField: "ORGANIZATION_ID",as: "organization"}},{$unwind: "$organization"},{$project: {_id: 0,organization_name: "$organization.NAME"}}])

            [Q]: What is the base cost of medication with the code 205923.
            [MongoDB]: db.patients.aggregate([    { $match: {"ENCOUNTERS.MEDICATIONS.CODE": 205923} },    { $unwind: "$ENCOUNTERS" },    { $unwind: "$ENCOUNTERS.MEDICATIONS" },    { $match: {"ENCOUNTERS.MEDICATIONS.CODE": 205923} },    { $project: { _id: 0, base_cost: "$ENCOUNTERS.MEDICATIONS.BASE_COST" } }])

            [Q]: What is the procedure code of the claim transaction 210ae4cd-7ca0-7da4-66a7-ef20b4f5db4d?
            [MongoDB]: db.patients.aggregate([    {        $match: {            "CLAIMS.CLAIM_TRANSACTIONS.CLAIM_TRANSACTION_ID": "210ae4cd-7ca0-7da4-66a7-ef20b4f5db4d"        }    },    {        $unwind: "$CLAIMS"    },    {        $unwind: "$CLAIMS.CLAIM_TRANSACTIONS"    },    {        $match: {            "CLAIMS.CLAIM_TRANSACTIONS.CLAIM_TRANSACTION_ID": "210ae4cd-7ca0-7da4-66a7-ef20b4f5db4d"        }    },    {        $project: {            _id: 0,             procedure_code: "$CLAIMS.CLAIM_TRANSACTIONS.PROCEDURE_CODE"        }    }]);


            With all the information given, provide a MongoDB query to the following question:

            [Q]: '{question}'
            [MongoDB]: 
\end{lstlisting}

\subsection{Database Schemas}
\label{app:db-schemas}

In the following pages, we provide the schemas for our four database models in a visual form.

\begin{figure}[h]
    \centering
    \includegraphics[height=1.6\textwidth]{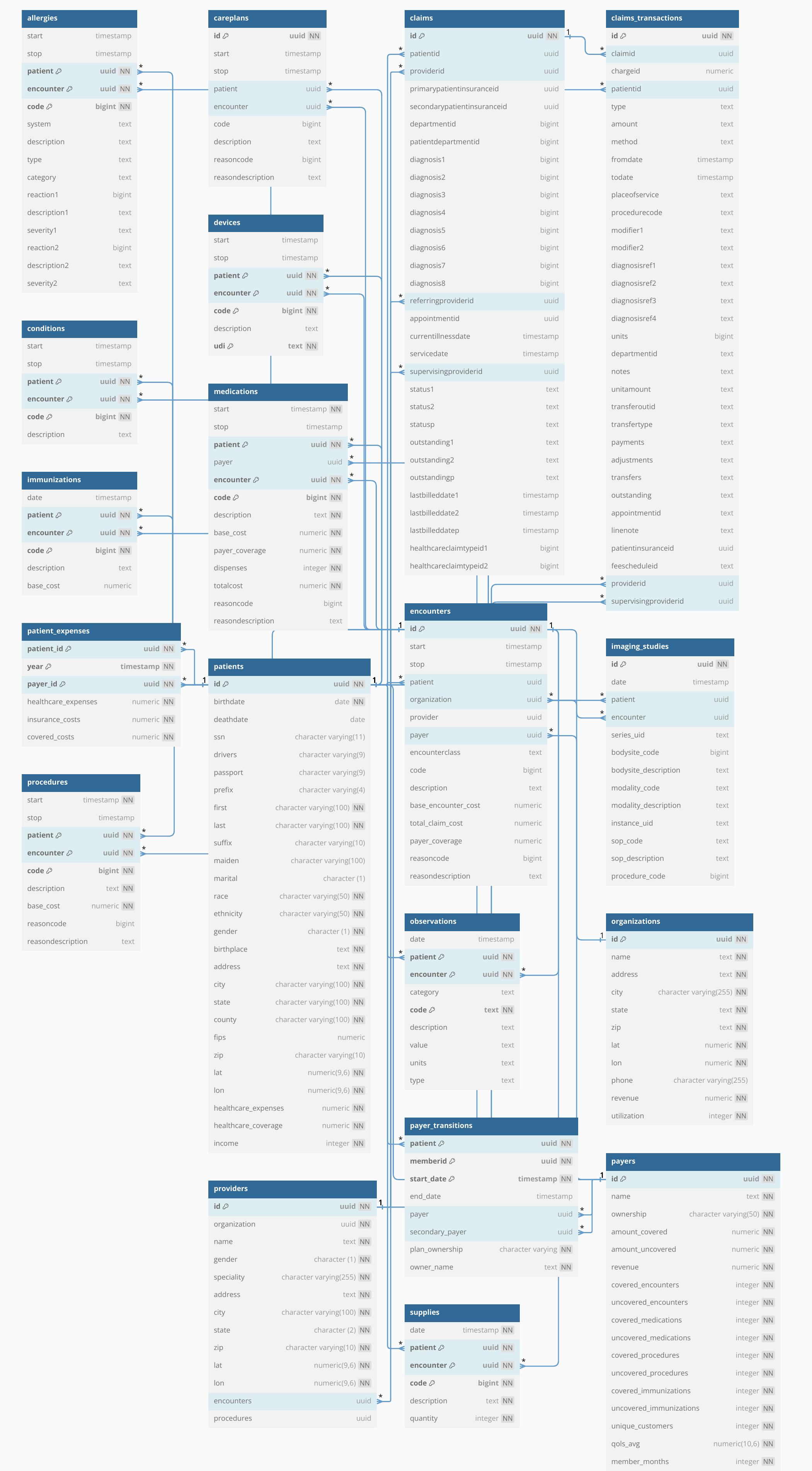}
    \caption{PostgreSQL database schema.}
    \label{fig:postgres_schema}
\end{figure}


\begin{figure}[h]
    \centering
    \includegraphics[width=0.9\textwidth]{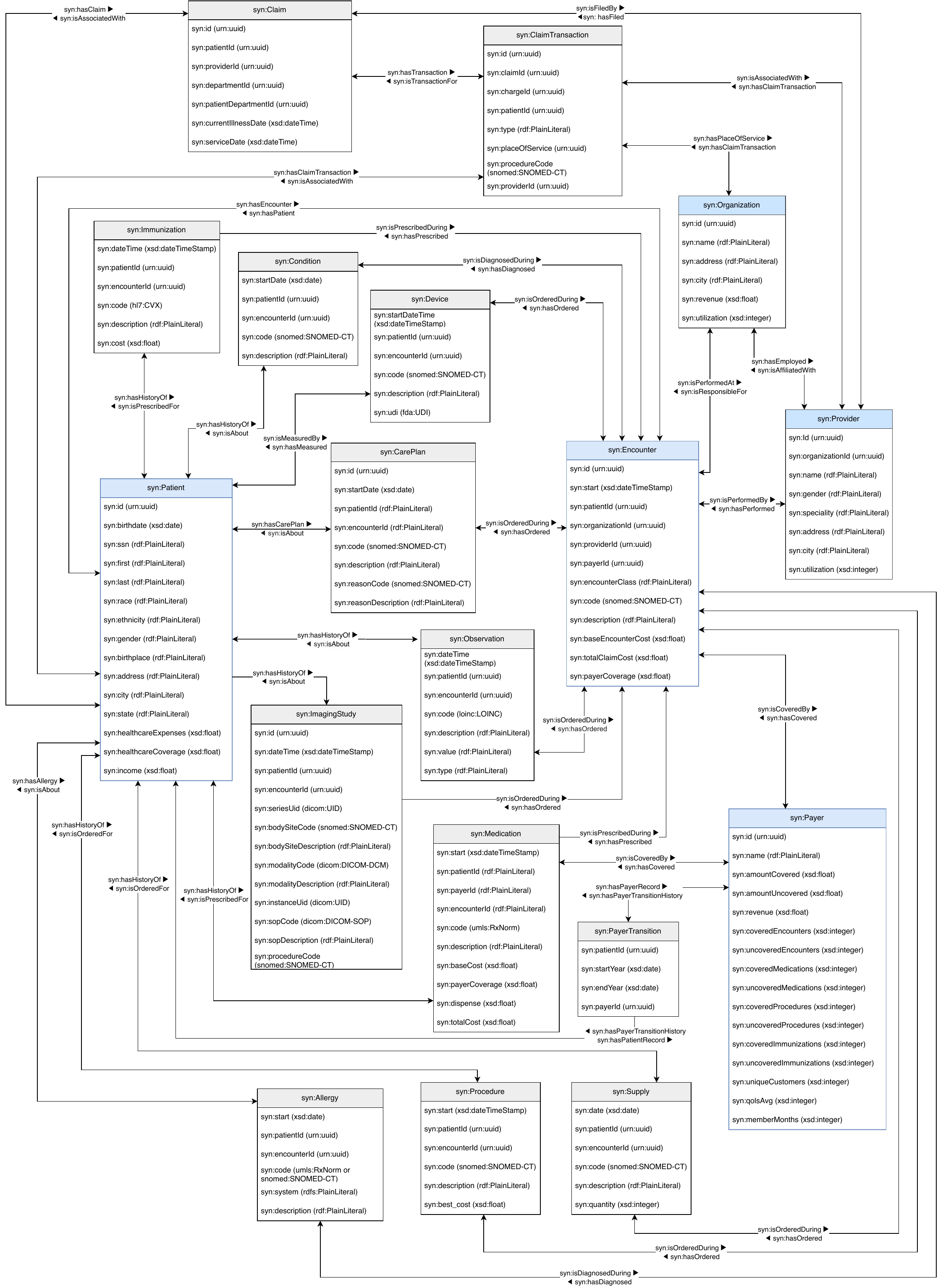}
    \caption{SPARQL UML ontology.}
    \label{fig:synthea_ontology}
\end{figure}


\begin{figure}[h]
    \includegraphics[width=1\textwidth]{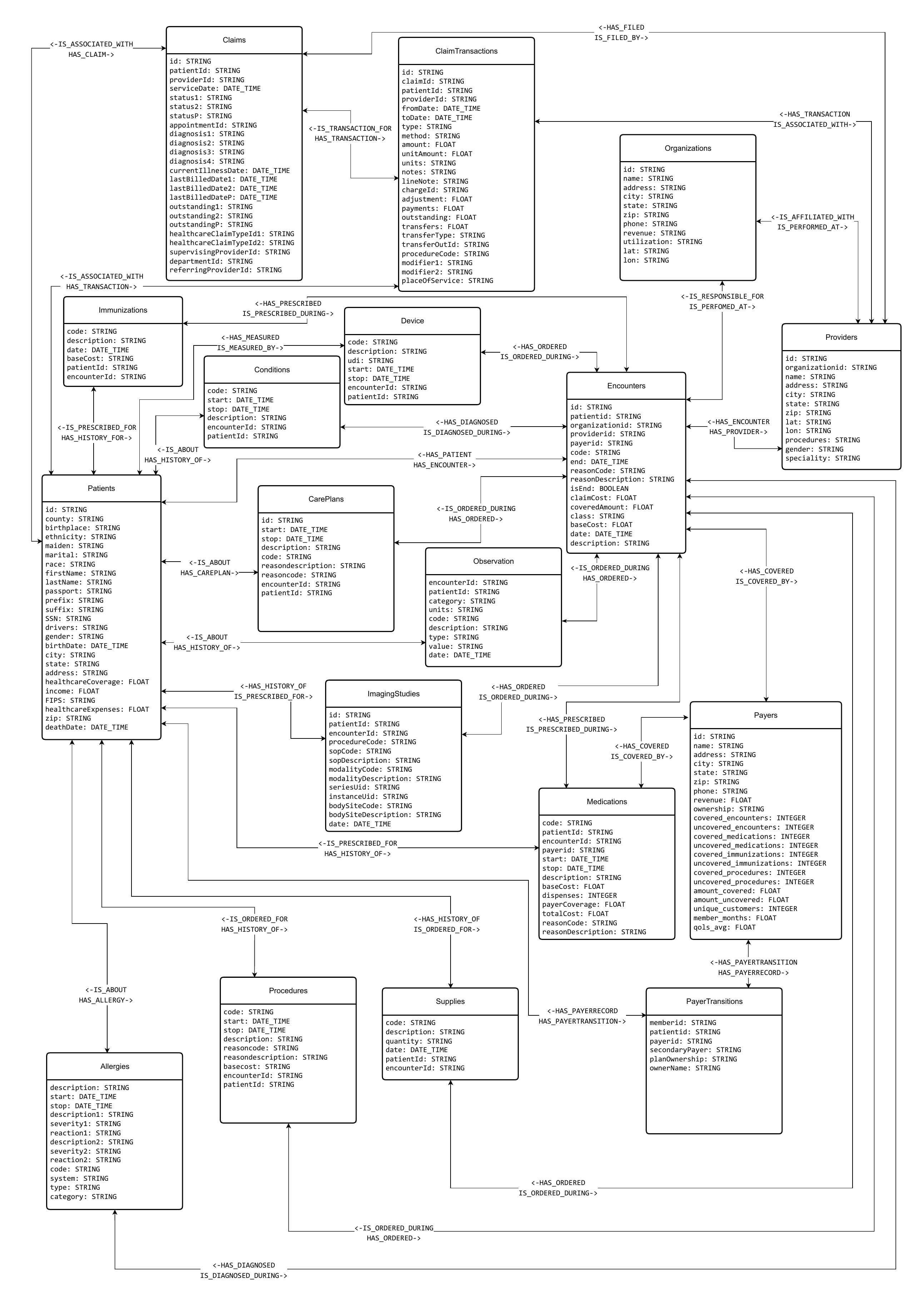}
    \caption{Cypher UML schema.}
    \label{fig:cypher_schema}
\end{figure}

\begin{figure}[h]
    \includegraphics[height=1\textheight, trim={0 0 0 10},clip]{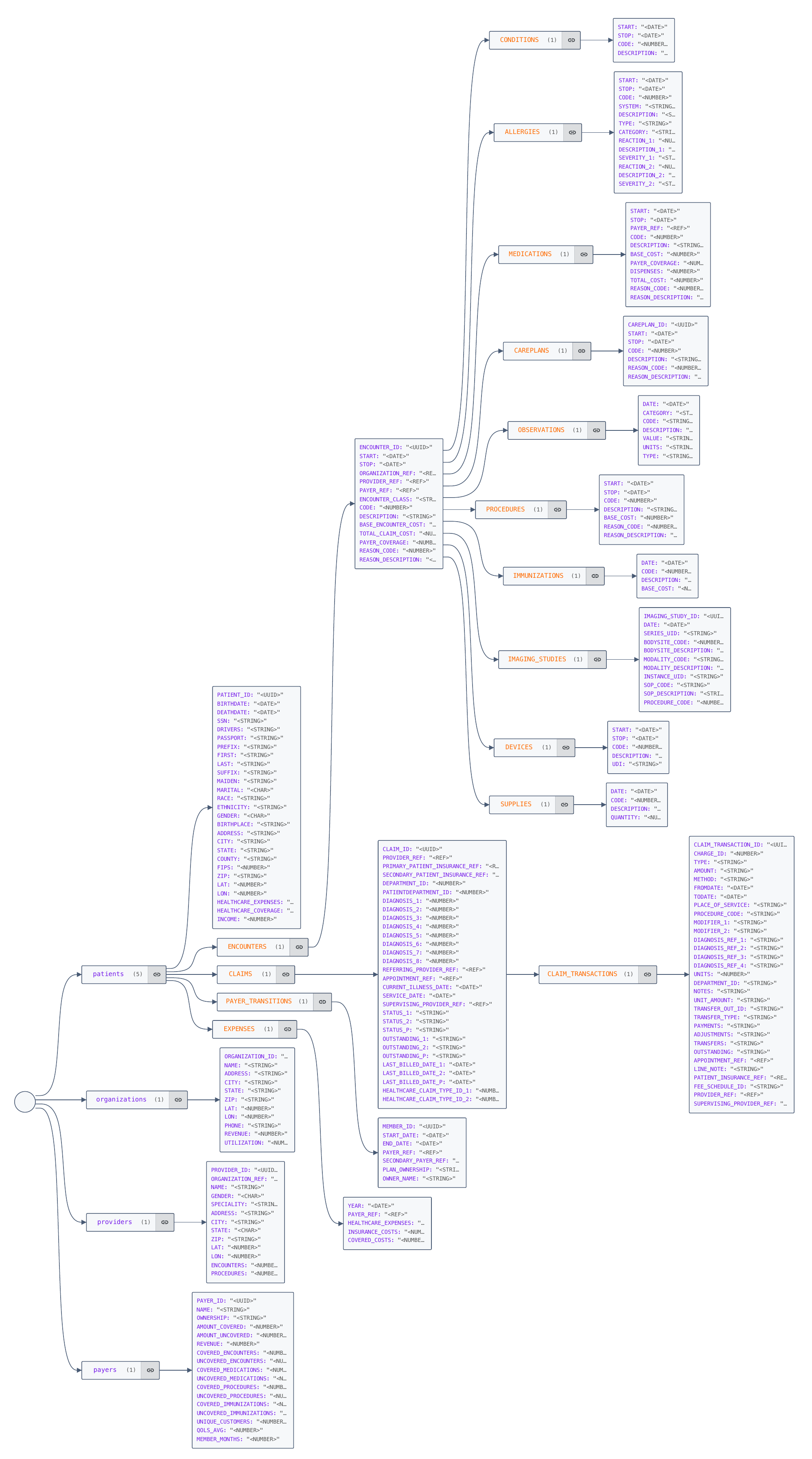}
    \caption{MongoDB schema (tree form).}
    \label{fig:mongodb_schema}
\end{figure}

\end{document}